\documentclass[11pt,a4paper]{article}
\pdfoutput=1
\usepackage{jheppub}
\usepackage{amsmath}
\usepackage{amssymb}
\usepackage{graphics}
\usepackage[active]{srcltx}
\usepackage{pdfsync}
\usepackage{shuffle}
\usepackage{slashed}

\usepackage{subfigure}

\newcommand{\bea}{\begin{eqnarray}}
\newcommand{\eea}{\end{eqnarray}}
\newcommand{\nn}{\nonumber \\}

\def\W #1{\widetilde{#1}}

\def\Tr{\mathop{\rm Tr}}

\def\eref#1{(\ref{#1})}

\def\b{{\beta}}

\def\Label#1{\label{#1}%
  \smash{\hbox to0pt{\raise1ex\hbox{\tiny[#1]}\hss}}}

\title{Tree level amplitudes from soft theorems}
\author[a]{Kang Zhou,}

\affiliation[a]{Center for Gravitation and Cosmology, College of Physical Science and Technology, Yangzhou University,\\
No.180, Siwangting Road, Yangzhou, 225009, P.R. China.}

\emailAdd{zhoukang@yzu.edu.cn}

\date{\today}
\abstract{We demonstrate that the tree level amplitudes and the explicit formulas of soft factors can be uniquely determined by soft theorems and the universality of soft factors. By imposing the soft theorems and the universality, as well as the assumption of double copy, we reconstruct single trace Yang-Mills-scalar amplitudes and pure Yang-Mills amplitudes, in the expanded formulas. The explicit formulas of soft factors for the bi-adjoint scalar and gluon are also determined. The expansions of Yang-Mills-scalar and Yang-Mills amplitudes can be extended to Einstein-Yang-Mills and gravitational amplitudes, and we use the expanded single trace Einstein-Yang-Mills amplitudes to reproduce the soft factors for the graviton.
}


\begin{document}

\maketitle \flushbottom

\section{Introduction}
\label{secintro}

Among the investigations of scattering amplitudes in the past decade, one of the remarkable progresses is the study of soft theorems.
Historically, soft theorems at tree level were derived using Feynman rules, originally discovered for photons in \cite{Low}, and extended to gravitons
in \cite{Weinberg}. In 2014, soft theorems have been revived for gravity (GR) \cite{Cachazo:2014fwa} and Yang-Mills (YM) theory \cite{Casali:2014xpa} for tree level amplitudes, by applying Britto-Cachazo-Feng-Witten (BCFW) recursion relation \cite{Britto:2004ap,Britto:2005fq}. For GR, the soft theorem was generalized from the leading order to sub-leading and sub-sub-leading orders.
For YM theory, the soft theorem was proven at the leading and sub-leading orders. These new results were generalized to arbitrary space-time dimensions \cite{Schwab:2014xua,Afkhami-Jeddi:2014fia}, via Cachazo-He-Yuan (CHY) formulas \cite{Cachazo:2013gna,Cachazo:2013hca, Cachazo:2013iea, Cachazo:2014nsa,Cachazo:2014xea}. One of the motivations, which have caught researchers' attention for studying soft theorems, is their equivalence to memory effects and asymptotic symmetries \cite{Strominger:2013jfa,Strominger:2013lka,He:2014laa,Kapec:2014opa,Strominger:2014pwa,Pasterski:2015tva,Barnich1,Barnich2,Barnich3}. Subsequently, the soft theorems and asymptotic symmetries for a variety of other theories including string theory, and the soft theorems at the loop level, have been further investigated in \cite{ZviScott,HHW,FreddyEllis,Bianchi:2014gla,Campiglia:2014yka,Campiglia:2016efb,Elvang:2016qvq,Guerrieri:2017ujb,Hamada:2017atr,Mao:2017tey,
Li:2017fsb,DiVecchia:2017gfi,
Bianchi:2016viy,Chakrabarti:2017ltl,Sen:2017nim,Hamada:2018vrw}. Meanwhile, the soft theorems were exploited in the construction of tree level amplitudes, such as using another type of soft behavior called the Adler zero to construct amplitudes of various effective theories, and the inverse soft theorem program, and so on \cite{Cheung:2014dqa,Luo:2015tat,Elvang:2018dco,Cachazo:2016njl,Rodina:2018pcb,Boucher-Veronneau:2011rwd,Nguyen:2009jk}.

Soft theorems describe the universal behavior of scattering amplitudes when one or more external massless momenta are taken to near zero. This limit can be achieved by re-scaling
the massless momenta via a soft parameter as $k^\mu\to \tau k^\mu$, and taking the limit $\tau\to 0$. Soft theorems then state the factorization
of the amplitudes. For instance, the $n+1$-point GR amplitude factorizes as
\bea
{\cal A}_{n+1}\,\to\,\Big(\tau^{-1}\,S^{(0)}_h+\tau^0\,S^{(1)}_h+\tau\,S^{(2)}_h+\cdots\Big)\,{\cal A}_n\,,~~~~\label{softtheo}
\eea
where ${\cal A}_n$ is the sub-amplitude of ${\cal A}_{n+1}$, which is generated from ${\cal A}_{n+1}$ by removing the soft external graviton.
The universal operators $S^{(0)}_h$, $S^{(1)}_h$, $S^{(2)}_h$ are called soft factors, or soft operators, at leading, sub-leading, and sub-sub-leading orders. The factorization in \eref{softtheo} is intuitively natural. Roughly speaking, in the soft limit the soft particle can be thought as vanishing, leaving a lower-point amplitude with the soft external leg removed, and the universal soft factors carried by the soft particle. Due to this clear physical picture for soft theorems,
it is interesting to ask, suppose we regard soft theorems as the basic principle, insist the universality of soft factors without knowing the explicit formulas of them,
what constraints will be imposed on amplitudes, and soft factors themselves? This is the main motivation for the current paper.

In this paper, by imposing the soft theorems and the universality of soft factors, and assuming the double copy structure \cite{Kawai:1985xq,Bern:2008qj,Chiodaroli:2014xia,Johansson:2015oia,Johansson:2019dnu}, we reconstruct single trace Yang-Mills-scalar (YMS) tree amplitudes and pure YM tree amplitudes, in the formulas
of expanding these amplitudes to double color ordered bi-adjoint scalar (BAS) amplitudes, established in \cite{Fu:2017uzt,Teng:2017tbo,Du:2017kpo,Du:2017gnh,Feng:2019tvb}. The leading soft factor
for the BAS scalar, the leading and sub-leading soft factors for the gluon, are also determined. Through the double copy structure , the expansions of YMS and YM amplitudes can be extended to expansions of single trace Einstein-Yang-Mills (EYM) amplitudes and the pure GR amplitudes. Then, by using the expanded formulas of EYM amplitudes, we reproduce the
soft factors for the graviton at leading, sub-leading, and sub-sub-leading orders.

It is worth to clarify some conventions which will be used in the remainder of this paper. First, when saying soft theorems, we mean the universal factorizations as in \eref{softtheo} (maybe only contains leading order, or leading and sub-leading orders), without knowing the formulas of soft factors. In other words, we insist the factorization property, regard it as a principle. Secondly, for latter convenience, from now on we absorb $\tau^a$ in \eref{softtheo} into the soft factors. Thirdly, the soft factor may includes kinematic variables carried by other hard particles, such as momenta and polarization vectors.
We say the soft operator acts on an external particle if this particle contributes the kinematic variables to the soft operator, and say the soft operator does not act on an external particle oppositely. As can be seen from our results, soft factors act on external particles which carry appropriate charges. Finally, when saying the universality of the soft operator, we mean it acts on external particles in universal manners. These manners of actions depend on the types of external particles, as will be further discussed in subsection.\ref{subsecBAS} and subsection.\ref{subsecYS1g}.

The remainder of this paper is organized as follows. In section.\ref{secreview}, we give a brief introduction to the BAS tree amplitudes, and the expansions of other amplitudes to them. In section.\ref{secYS}, we use the soft theorems and the universality of soft factors, to construct the single trace YMS tree amplitudes in the expanded formulas, as well as the soft factors for the scalar and gluon. In section.\ref{secYM}, we use the similar technic to
determine the pure YM tree amplitudes in the expanded formulas. Then, in section.\ref{secEYMGR}, we extend the expanded formulas to single trace EYM tree amplitudes and GR tree amplitudes, and figure out the soft factors for the graviton. Finally, we close with conclusion and discussion in section.\ref{secconclu}.

\section{Brief review of BAS theory and expansions of amplitudes}
\label{secreview}

For readers' convenience, in this section we rapidly review the necessary background. In subsection.\ref{subsecBAS}, we review the tree level amplitudes of bi-adjoint scalar (BAS) theory. Some notations and technics which will be used in subsequent sections are also introduced. In subsection.\ref{subsecexpand}, we discuss the expansions of tree amplitudes to BAS amplitudes, including the choice of basis, as well as the double copy structure for coefficients.

\subsection{Tree level BAS amplitudes}
\label{subsecBAS}

The BAS theory describes the bi-adjoint scalar field $\phi_{a\bar{a}}$ with the Lagrangian
\bea
{\cal L}_{BAS}={1\over2}\,\partial_\mu\phi^{a\bar{a}}\,\partial^{\mu}\phi^{a\bar{a}}-{\lambda\over3!}\,f^{abc}f^{\bar{a}\bar{b}\bar{c}}\,
\phi^{a\bar{a}}\phi^{b\bar{b}}\phi^{c\bar{c}}\,,
\eea
where the structure constant $f^{abc}$ and generator $T^a$ satisfy
\bea
[T^a,T^b]=if^{abc}T^c\,,
\eea
and the dual color algebra encoded by $f^{\bar{a}\bar{b}\bar{c}}$ and $T^{\bar{a}}$ is analogous.
The tree level amplitudes of this theory contain only propagators, and can be decomposed into double color ordered partial amplitudes via the standard technic.
Each double color ordered partial amplitude is simultaneously planar with respect to two color orderings, arise from expanding the full $n$-point amplitude to $\Tr(T^{a_{\sigma_1}}\cdots T^{a_{\sigma_n}})$ and $\Tr(T^{\bar{a}_{\bar{\sigma}_1}}\cdots T^{\bar{a}_{\bar{\sigma}_n}})$ respectively,
where $\sigma_i$ and $\bar{\sigma}_i$ denote permutations among all external scalars.

\begin{figure}
  \centering
  \includegraphics[width=6cm]{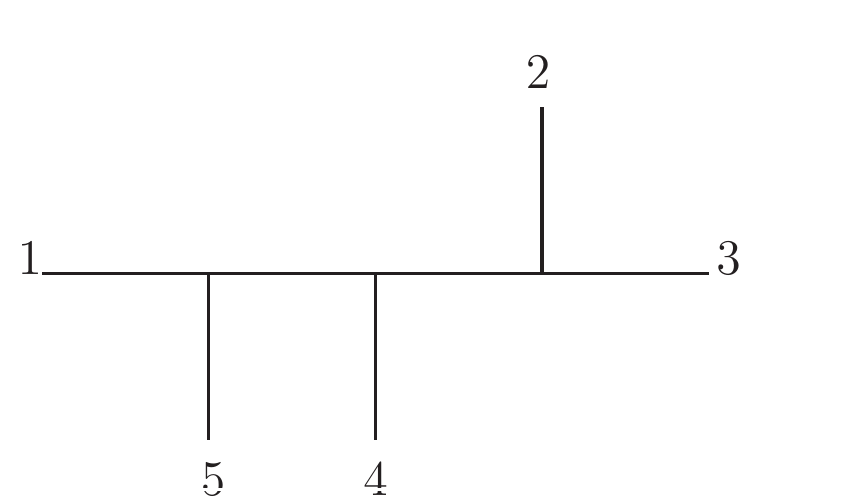}
   \includegraphics[width=6cm]{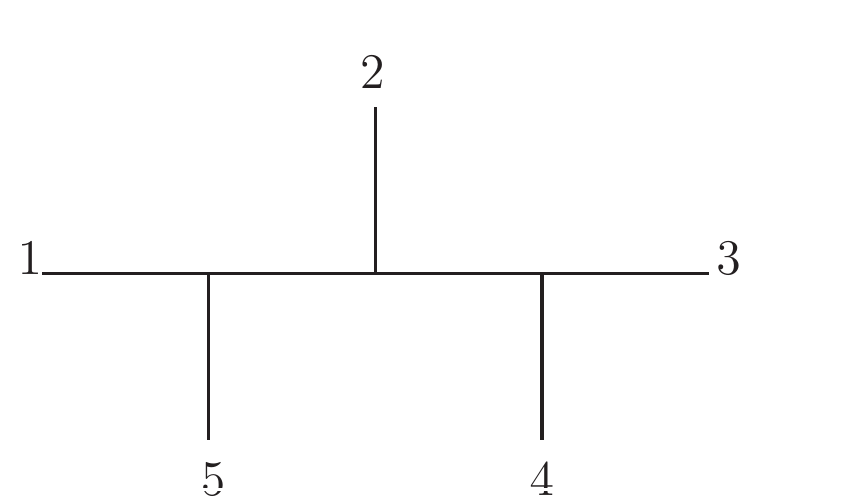}  \\
  \caption{Two $5$-point diagrams}\label{5p}
\end{figure}

\begin{figure}
  \centering
   \includegraphics[width=4cm]{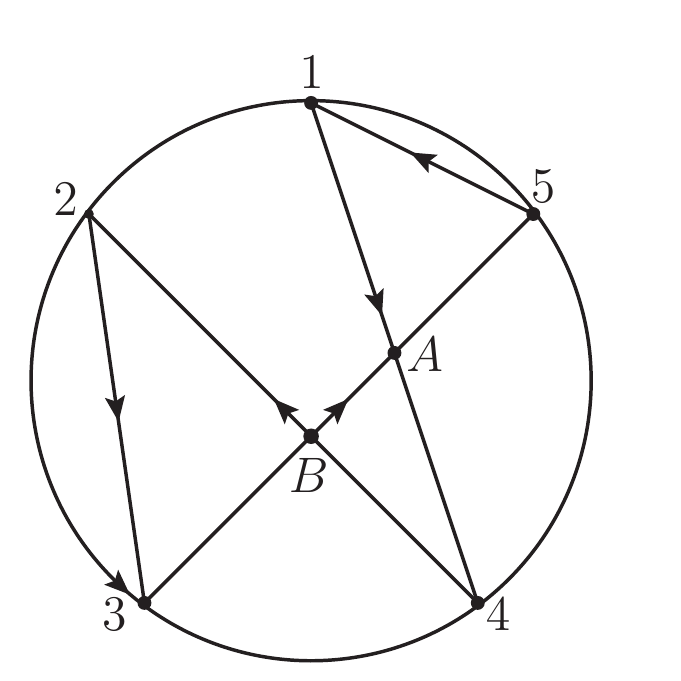}
   \includegraphics[width=4cm]{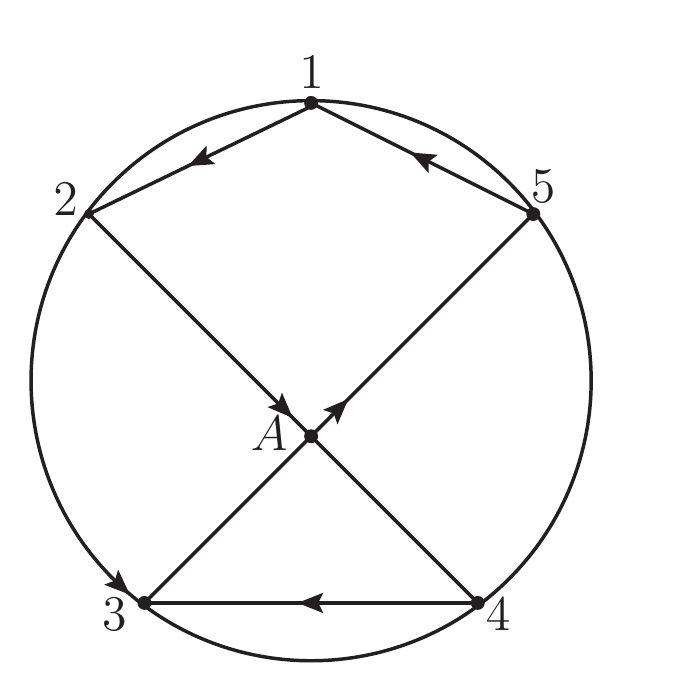} \\
  \caption{Diagram for ${\cal A}_S(1,2,3,4,5|1,4,2,3,5)$ and ${\cal A}_S(1,2,3,4,5|1,2,4,3,5)$.}\label{dis14235}
\end{figure}

To calculate double color ordered partial amplitudes, it is convenient to employ the diagrammatical method proposed by Cachazo, He and Yuan in \cite{Cachazo:2013iea}.
For a BAS amplitude whose double color-orderings are given, this method provides the corresponding Feynman diagrams as well as the overall sign directly.
Let us consider the $5$-point example ${\cal A}_S(1,2,3,4,5|1,4,2,3,5)$.
In Figure.\ref{5p}, the first diagram satisfies both two color orderings $(1,2,3,4,5)$ and $(1,4,2,3,5)$, while the second one satisfies the ordering
$(1,2,3,4,5)$ but not $(1,4,2,3,5)$. Thus, the first diagram is allowed by the double color orderings $(1,2,3,4,5|1,4,2,3,5)$, while the second one is not. It is easy to see other diagrams are also forbidden by the ordering $(1,4,2,3,5)$, thus the first diagram in Figure.\ref{5p} is the only diagram contributes to the amplitude
${\cal A}_S(1,2,3,4,5|1,4,2,3,5)$.
The Feynman diagrams contribute to a given BAS amplitude can be obtained via systematic diagrammatical rules. For the above example, one can draw a disk diagram as follows.
\begin{itemize}
\item Draw points on the boundary of the disk according to the first ordering $(1,2,3,4,5)$.
\item Draw a loop of line segments which connecting the points according to the second ordering $(1,4,2,3,5)$.
\end{itemize}
The obtained disk diagram is shown in the first diagram in Figure.\ref{dis14235}. From the diagram, one can see that two orderings share the boundaries $\{1,5\}$ and $\{2,3\}$. These co-boundaries
indicate channels ${1/s_{15}}$ and ${1/s_{23}}$, therefore the first Feynman diagram in Figure.\ref{5p}. Then the BAS amplitude ${\cal A}_S(1,2,3,4,5|1,4,2,3,5)$ can be computed as
\bea
{\cal A}_S(1,2,3,4,5|1,4,2,3,5)={1\over s_{23}}{1\over s_{51}}\,,
\eea
up to an overall sign. The Mandelstam variable $s_{i\cdots j}$ is defined as
\bea
s_{i\cdots j}\equiv K_{ij}^2\,,~~~~K_{ij}\equiv\sum_{a=i}^j\,k_a\,,~~~~\label{mandelstam}
\eea
where $k_a$ is the momentum carried by the external leg $a$.

As another example, let us consider the BAS amplitude ${\cal A}_S(1,2,3,4,5|1,2,4,3,5)$. The corresponding disk diagram is shown in the second configuration in
Figure.\ref{dis14235}, and one can see two orderings have co-boundaries $\{3,4\}$ and $\{5,1,2\}$. The co-boundary $\{3,4\}$ indicates the channel ${1/ s_{34}}$. The co-boundary $\{5,1,2\}$ indicates the channel ${1/s_{512}}$ which is equivalent to $1/s_{34}$, as well as sub-channels ${1/ s_{12}}$ and ${1/ s_{51}}$. Using the above decomposition, one can calculate ${\cal A}_S(1,2,3,4,5|1,2,4,3,5)$ as
\bea
{\cal A}_S(1,2,3,4,5|1,2,4,3,5)={1\over s_{34}}\Big({1\over s_{12}}+{1\over s_{51}}\Big)\,,
\eea
up to an overall sign.

The overall sign, determined by the color algebra, can be fixed by following rules.
\begin{itemize}
\item Each polygon with odd number of vertices contributes
a plus sign if its orientation is the same as that of the disk and a minus sign if opposite.
\item Each polygon with even number of vertices always contributes a minus sign.
\item Each intersection point contributes a minus sign.
\end{itemize}
We now apply these rules to previous examples. In the first diagram in Figure.\ref{dis14235}, the polygons are three triangles, namely $51A$, $A4B$ and $B23$, which contribute $+$, $-$, $+$ respectively, while two intersection points $A$ and $B$ contribute two $-$. In the second one in Figure.\ref{dis14235}, the polygons are $512A$ and $A43$, which contribute two $-$, while the intersection point $A$ contributes $-$. Then we arrive at the full results
\bea
{\cal A}_S(1,2,3,4,5|1,4,2,3,5)&=&-{1\over s_{23}}{1\over s_{51}}\,,\nn
{\cal A}_S(1,2,3,4,5|1,2,4,3,5)&=&-{1\over s_{34}}\Big({1\over s_{12}}+{1\over s_{51}}\Big)\,.
\eea

\begin{figure}
  \centering
   \includegraphics[width=4cm]{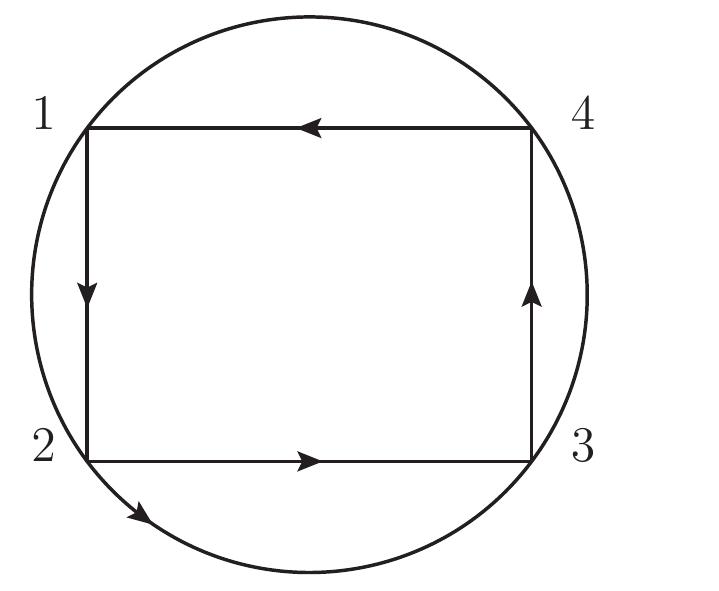} \\
  \caption{The overall sign $+$ under the new convention.}\label{newconvention}
\end{figure}

In the reminder of this paper, we adopt another convention for the overall sign. If the line segments form a convex polygon, and the orientation of the convex polygon is the same as that of the disk, then the overall sign is $+$. For instance, the disk diagram in Figure.\ref{newconvention} indicates the overall sign $+$ under the new convention, while the old convention gives $-$ according to the square formed by four line segments. Notice that the diagrammatical rules described previously still give the related sign between different disk diagrams. For example, two disk diagrams in Figure.\ref{dis14235} shows that the relative sign between ${\cal A}_S(1,2,3,4,5|1,4,2,3,5)$ and
${\cal A}_S(1,2,3,4,5|1,2,4,3,5)$ is $+$.

\begin{figure}
  \centering
   \includegraphics[width=8cm]{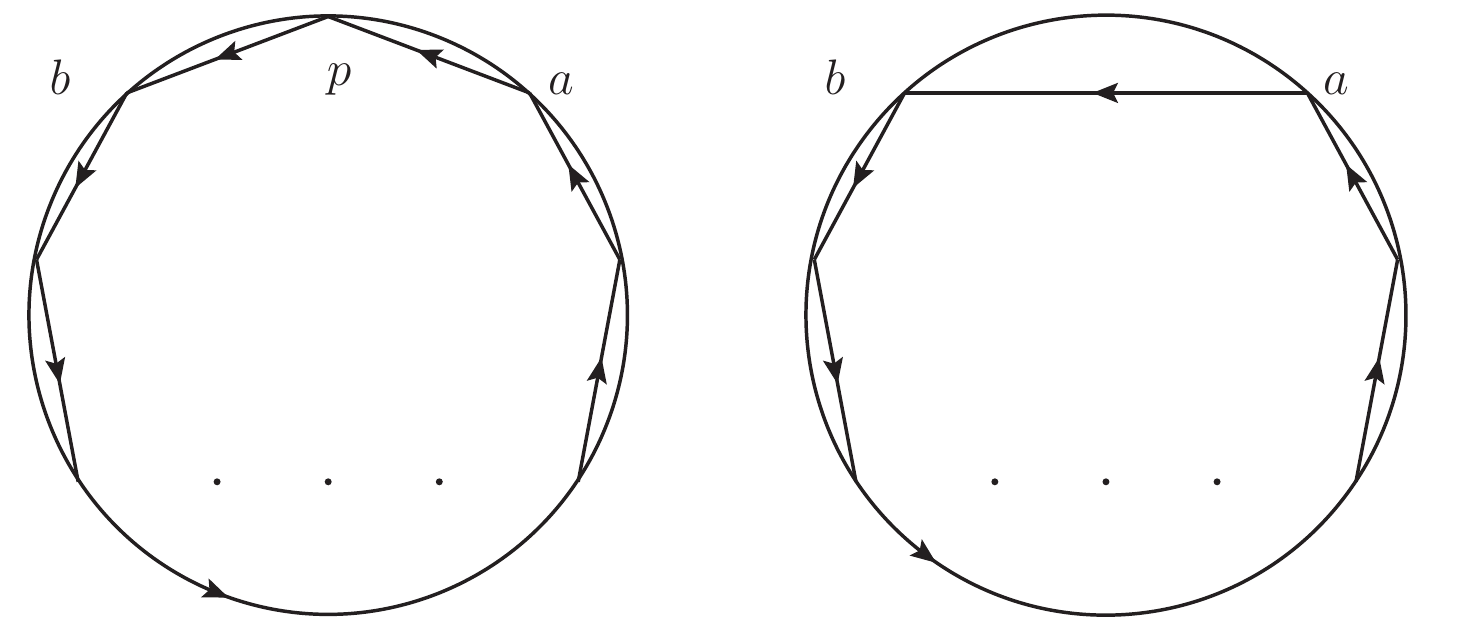} \\
  \caption{Two convex polygons correspond to ${\cal A}_S(\cdots,a,p,b,\cdots|\cdots,a,p,b,\cdots)$ and
${\cal A}_S(\cdots,a,b,\cdots|\cdots,a,b,\cdots)$.}\label{polygon}
\end{figure}

\begin{figure}
  \centering
   \includegraphics[width=8cm]{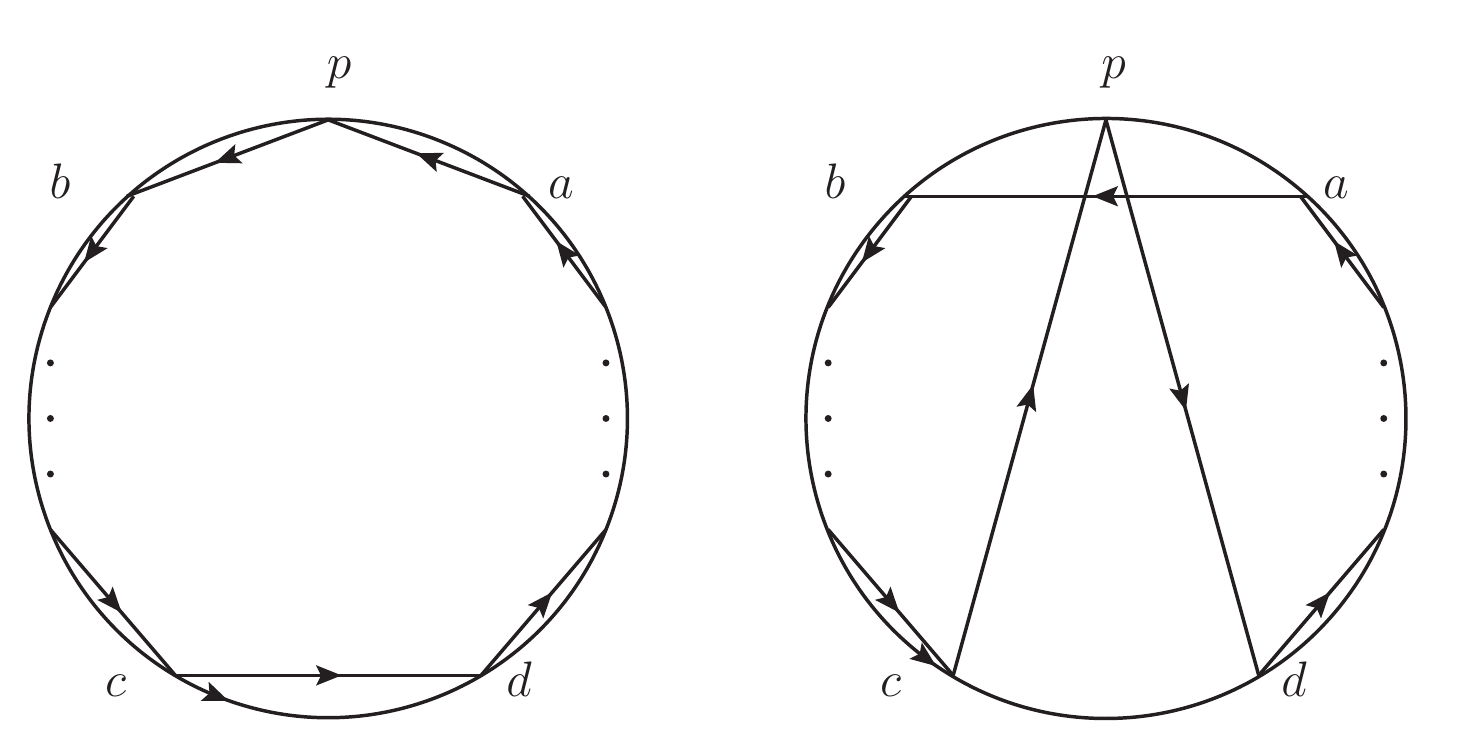} \\
  \caption{Disk diagrams for ${\cal A}_S(\cdots,a,p,b,\cdots|\cdots,a,p,b,\cdots)$ and
${\cal A}_S(\cdots,a,p,b,\cdots|\cdots,c,p,d,\cdots)$.}\label{poly1}
\end{figure}

One application of the new convention, which is crucial in subsequent sections, is as follows. For the double color ordered BAS amplitude ${\cal A}_S(\cdots,a,p,b,\cdots|\cdots,a,p,b,\cdots)$, suppose we remove the external scalar $p$, the overall sign for the resulted amplitude
${\cal A}_S(\cdots,a,b,\cdots|\cdots,a,b,\cdots)$ is $+$. Notice that in the notations above we do not require the full color orderings at the l.h.s and r.h.s of $|$ to be the same. On the other hand, if we remove $p$ from ${\cal A}_S(\cdots,a,p,b,\cdots,c,d,\cdots|\cdots,a,b,\cdots,c,p,d,\cdots)$, whose color orderings are the same as those for ${\cal A}_S(\cdots,a,p,b,\cdots|\cdots,a,p,b,\cdots)$ except the positions of $p$, the obtained amplitude ${\cal A}_S(\cdots,a,b,\cdots|\cdots,a,b,\cdots)$ carries an overall $-$. We now give the interpretation to the above observation.
One can change the color orderings for ${\cal A}_S(\cdots,a,p,b,\cdots|\cdots,a,p,b,\cdots)$ and
${\cal A}_S(\cdots,a,b,\cdots|\cdots,a,b,\cdots)$ simultaneously, to arrive at two convex polygons in disk diagrams in Figure.\ref{polygon}.
This procedure will not alter the relative sign. Each diagram in Figure.\ref{polygon} corresponds to an overall
$+$, due to our new convention. Thus, generating ${\cal A}_S(\cdots,a,b,\cdots|\cdots,a,b,\cdots)$ from ${\cal A}_S(\cdots,a,p,b,\cdots|\cdots,a,p,b,\cdots)$ will not create the overall $-$.
Then, we change the color orderings for ${\cal A}_S(\cdots,a,p,b,\cdots|\cdots,a,p,b,\cdots)$ and ${\cal A}_S(\cdots,a,p,b,\cdots,c,d,\cdots|\cdots,a,b,\cdots,c,p,d,\cdots)$ simultaneously to get two configurations in Figure.\ref{poly1},
where the line segments in first one again form a convex polygon. The relative sign between these two configurations in Figure.\ref{poly1} can be figured out via the diagrammatical rules. Suppose the number of points on the boundary is odd, the old convention indicates that the polygon carries a $+$. In the second configuration, two intersection points contribute two $-$, the triangle carries the orientation opposite to the disk thus contributes $-$, and two convex polygons whose numbers of vertices are both odd or both even contribute $+$. Consequently, there is a relative $-$ between two configurations. If the number of points on the boundary is even, the first configuration carries $-$. The second configuration contains two convex polygons which carry the same orientations as the disk,
and the numbers of their vertices are even for one polygon and odd for another one, thus the total number of $-$ for the second configuration is $4$. Again, there is a relative $-$ between two configurations. For the first configuration, the new convention for overall sign implies that removing $p$  will not create $-$. Thus, for the second one, removing $p$ provides a $-$ to the resulted amplitude.

Then, we discuss which propagators can contribute to a double color ordered BAS amplitude with fixed color orderings.
In the diagrammatical rules, this question is answered by the co-boundaries of the disk diagram. One can also deduce the following new
rule which will be used latter. For the set $\alpha$ of successive points on the boundary of disk, let us call
line segments connecting two points in $\alpha$ as internal lines, and line segments connecting one point in $\alpha$ and another one in $\bar{\alpha}$ as external lines, where $\bar{\alpha}$ is the complementary set of $\alpha$. The new rule is, the set $\alpha$ contributes the propagator $1/s_{\alpha}$ to the BAS partial amplitude if and only if it has two external lines. This rule is manifestly equivalent to requiring the co-boundary.

\begin{figure}
  \centering
   \includegraphics[width=8cm]{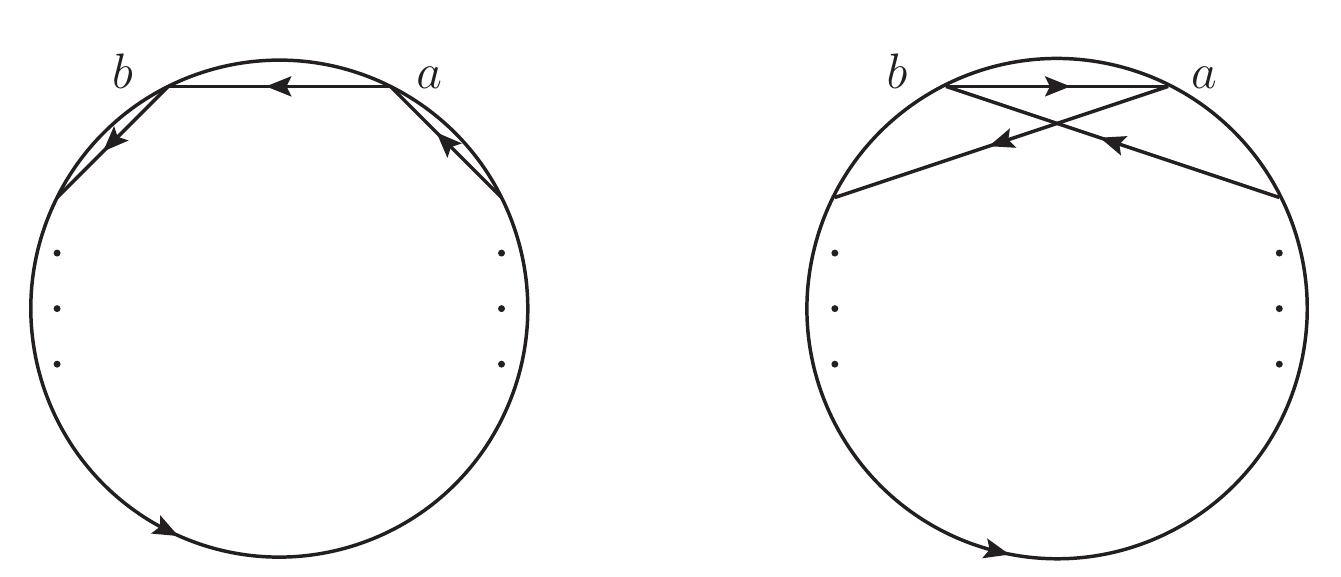} \\
  \caption{Disk diagrams for ${\cal A}_S(\cdots,a,p,b,\cdots|\cdots,a,p,b,\cdots)$ and
${\cal A}_S(\cdots,a,p,b,\cdots|\cdots,c,p,d,\cdots)$.}\label{delta}
\end{figure}

When considering the soft limit, the $2$-point channels play the central role. Since the partial BAS amplitude carries two color orderings, if the $2$-point channel contributes $1/s_{ab}$ to the amplitude, external legs $a$ and $b$ must be adjacent to each other in both two orderings.
Suppose the first color ordering is $(\cdots,a,b,\cdots)$, then $1/s_{ab}$ is allowed by this ordering. To denote if it is allowed by another one, we introduce the symbol $\delta_{ab}$ whose ordering of two subscripts $a$ and $b$ is determined by the first color ordering\footnote{The Kronecker symbol will not appear in this paper except in the third footnote, thus we hope the notation $\delta_{ab}$ will not confuse the readers.}. The value of $\delta_{ab}$ is $\delta_{ab}=1$ if another color ordering is $(\cdots,a,b,\cdots)$, $\delta_{ab}=-1$ if another color ordering is $(\cdots,b,a,\cdots)$, due to the ani-symmetry of the structure constant, i.e., $f^{abc}=-f^{bac}$, and $\delta_{ab}=0$
otherwise. The value of $\delta_{ab}$ is consistent with our new convention of the overall sign, as can be seen in Figure.\ref{delta}. The first configuration in Figure.\ref{delta} gives $\delta_{ab}=1$, due to the convention that the overall sign corresponds to a convex polygon is $+$. The second configuration carries a relative sign $-$ when comparing with the first one, as can be figured out via the diagrammatical rules, thus implies $\delta_{ab}=-1$. From the definition, it is straightforward to see $\delta_{ab}=-\delta_{ba}$. The symbol $\delta_{ab}$ will be used frequently latter.

Before ending this subsection, we determine the leading order soft factor for the BAS scalar. Consider the double color ordered BAS amplitude ${\cal A}_S(1,\cdots,n|\sigma)$, which carries two color orderings $(1,\cdots,n)$ and $\sigma$. We re-scale $k_i$
as $k_i\to\tau k_i$, and expand the amplitude in $\tau$. The leading order contribution manifestly aries from $2$-point channels $1/s_{1(i+1)}$
and $1/ s_{(i-1)i}$ which provide the $1/\tau$ order contributions, namely,
\bea
{\cal A}^{(0)}_S(1,\cdots,n|\sigma)&=&{1\over \tau}\Big({\delta_{i(i+1)}\over s_{i(i+1)}}+{\delta_{(i-1)i}\over s_{(i-1)i}}\Big)\,
{\cal A}_S(1,\cdots,i-1,\not{i},i+1,\cdots,n|\sigma\setminus i)\nn
&=&S^{(0)}_s(i)\,{\cal A}_S(1,\cdots,i-1,\not{i},i+1,\cdots,n|\sigma\setminus i)\,,~~~\label{for-soft-fac-s}
\eea
where $\not{i}$ stands for removing the leg $i$, $\sigma\setminus1$ means the color ordering generated from $\sigma$ by eliminating $i$. The leading soft operator $S^{(0)}_s(i)$ for the scalar $i$ is extracted as
\bea
S^{(0)}_s(i)={1\over \tau}\,\Big({\delta_{i(i+1)}\over s_{i(i+1)}}+{\delta_{(i-1)i}\over s_{(i-1)i}}\Big)\,,~~~~\label{soft-fac-s-0}
\eea
which acts on external scalars which are adjacent to $i$ in two color orderings.
This operator serves as one of starting points in subsection.\ref{subsecYS1g}.

It is worth to take \eref{soft-fac-s-0} as the instance, to give further explanations for the universality of soft operator.
The operator \eref{soft-fac-s-0} acts on external BAS scalars which are adjacent to $i$ in either of two color orderings, we assume such manner is universal.
For amplitudes contain other types of particles, such as scalars coupled to gluons, the leading soft factor $S^{(0)}_s(i)$ for scalar still acts on external
scalars in the manner described by \eref{soft-fac-s-0}, i.e., it always acts on external BAS scalars which are adjacent to $i$ in both of two color orderings. On the other hand, it is possible $S^{(0)}_s(i)$ also acts on other types of external particles such as gluons. This possibility can not be studied in the current case, since the pure BAS amplitudes only include external BAS scalars. One may wonder the
action of soft operator should depend on theories defined by Lagrangians, rather than types of external particles. For amplitudes under consideration in the current paper, this puzzle can be solved via the following argument. Let us take the single trace YMS amplitudes as the example. One can regard the pure BAS and pure YM amplitudes as special cases of general single trace YMS amplitudes\footnote{For the multiple trace YMS amplitudes, such point of view is not valid, since the BAS amplitude is single trace.}, thus the soft behavior of pure BAS amplitude when one external scalar being soft can be obtained by acting the soft operator for the single trace YMS amplitudes to the pure BAS one. The soft operator can be split into two parts, one acts on external scalars, another one acts on external gluons. The second part annihilates the pure BAS amplitude which does not include any external gluon, thus the effective part is only the first one. The first part must be equal to \eref{soft-fac-s-0}, otherwise the action of the first part can not reproduce the result \eref{for-soft-fac-s}. Therefore, for the single trace YMS case, the leading soft operator for the scalar always acts on external scalars
in the manner showed in \eref{soft-fac-s-0}, without regarding whether the amplitude contains external gluons. Similar discussion shows that the soft factors for the gluon for the single trace YMS amplitudes can be applied to pure YM ones, by removing the part which acts on external scalars.
One can also regard the YM and GR amplitudes as special cases of single trace EYM amplitudes and obtain the similar conclusion, the discussion is analogous.

\subsection{Expanding tree level amplitudes to BAS basis}
\label{subsecexpand}

\begin{figure}
  \centering
   \includegraphics[width=4cm]{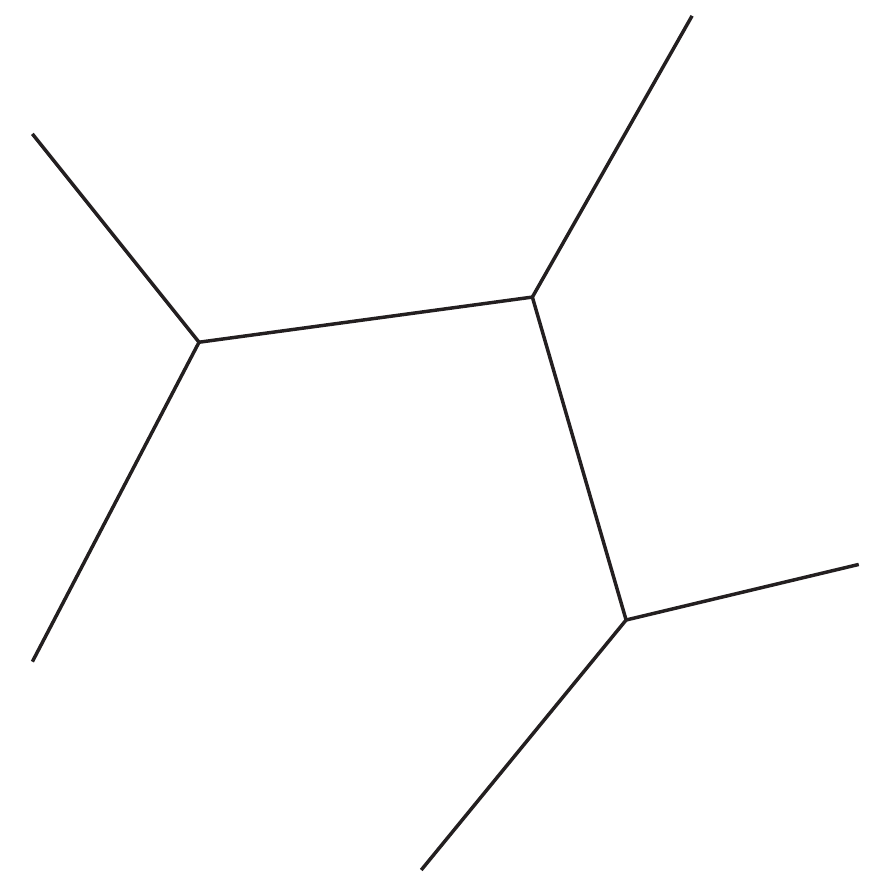}
   \includegraphics[width=4cm]{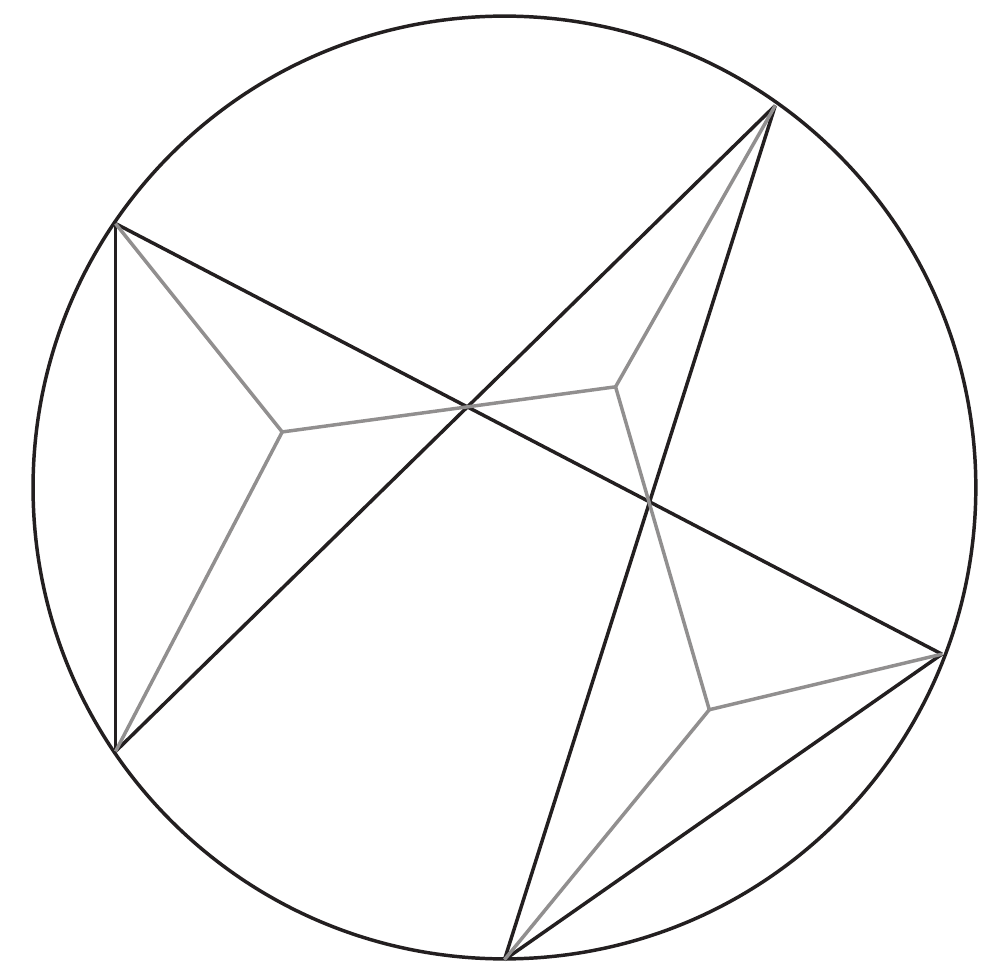} \\
  \caption{Map between Feynman diagram and disk diagram.}\label{example}
\end{figure}

Tree level amplitudes which contain only massless particles can be expanded to double color ordered BAS amplitudes,
due to the observation that each Feynman diagram for pure propagators can be mapped to at least one disk diagram whose polygons are all triangles. An illustrative example is given in Figure.\ref{example}. Since each tree amplitude can be expanded by tree Feynman diagrams, and each Feynman diagram contributes propagators together with a numerator without any pole, one can conclude that each tree amplitude can be expanded to double color ordered BAS amplitudes, with coefficients which are polynomials depend on Lorentz invariants created by external kinematical variables.
To realize the expansion, one need to find the basis consists of BAS amplitudes.
Such basis can be determined by the well known Kleiss-Kuijf (KK) relation \cite{Kleiss:1988ne}
\bea
{\cal A}_S(1,\alpha,n,\beta|\sigma)=(-)^{|\beta|}\,{\cal A}_S(1,\alpha\shuffle\beta^T,n|\sigma)\,,~~~\label{KK}
\eea
which is based on the color algebra. Here $\alpha$ and $\beta$ are two ordered subsets of external scalars, and $\beta^T$ stands for the ordered set generated from $\beta$ by reversing the original ordering. The BAS amplitude ${\cal A}_S(1,\alpha,n,\beta|\sigma)$ at the l.h.s of \eref{KK} carries two color orderings, one is $(1,\alpha,n,\beta)$, another one is denoted by $\sigma$. The symbol $\shuffle$ means summing over all
possible shuffles of two ordered subsets $\b_1$ and $\b_2$, i.e., all permutations in the set $\b_1\cup \b_2$ while preserving the orderings
of $\b_1$ and $\b_2$. For instance, suppose $\b_1=\{1,2\}$ and $\b_2=\{3,4\}$, then
\bea
\b_1\shuffle \b_2=\{1,2,3,4\}+\{1,3,2,4\}+\{1,3,4,2\}+\{3,1,2,4\}+\{3,1,4,2\}+\{3,4,1,2\}\,.~~~~\label{shuffle}
\eea
The analogous KK relation holds for another color ordering $\sigma$.
The KK relation implies that different double color ordered BAS amplitudes are not independent, thus the basis can be chosen as BAS amplitudes
${\cal A}_S(1,\sigma_1,n|1,\sigma_2,n)$, with $1$ and $n$ are fixed at two ends in each color ordering. We call such basis the KK BAS basis. Based on the discussion above, the KK BAS basis can provide any structure of propagators, thus any amplitude can be expanded to this basis, with coefficients which contain no pole\footnote{The well known Bern-Carrasco-Johansson (BCJ) relation \cite{Bern:2008qj,Chiodaroli:2014xia,Johansson:2015oia,Johansson:2019dnu} implies the relations among BAS amplitudes in the KK basis, and the independent BAS amplitudes can be obtained by fixing three legs at three particular positions in the color orderings. However, in the BCJ relation, coefficients of BAS amplitudes depend on Mandelstam variables, this character leads to poles in coefficients when expanding to BCJ basis. We hope all poles are contributed by propagators in amplitudes, and the coefficients contain no pole, thus chose the KK basis.}.

In this paper, we will consider the expansions of single trace YMS amplitudes, YM amplitudes, single trace EYM amplitudes, and GR amplitudes. We now discuss them one by one. In the single trace YMS amplitude ${\cal A}_{YS}(1,\sigma_1,n;p_1,\cdots,p_m|1,\sigma_2,n)$, the external scalars encoded by $i\in\{1,\cdots,n\}$ are included in two color orderings, one is
$(1,\sigma_1,n)$ among only scalars, another one is $(1,\sigma_2,n)$ among all external legs. The external gluons labeled by $p_j$ with $j\in\{1,\cdots,m\}$ belong to the color ordering $(1,\sigma_2,n)$, while $p_1,\cdots,p_m$ at the l.h.s of $|$ and r.h.s of $;$ are un-ordered. Here we fixed $1$ and $n$ at two ends in each color ordering, due to the KK relation. The amplitude ${\cal A}_{YS}(1,\sigma_1,n;p_1,\cdots,p_m|1,\sigma_2,n)$ can be expanded to KK BAS basis as
\bea
{\cal A}_{YS}(1,\sigma_1,n;p_1,\cdots,p_m|1,\sigma_2,n)=\sum_{\sigma_3}\,{\cal C}(\sigma_1,\sigma_3,\epsilon_i,k_i)\,{\cal A}_{S}(1,\sigma_3,n|1,\sigma_2,n)\,,~~~\label{exp-YS-KK}
\eea
where $\sigma_3$ are permutations among external legs in $\{2,\cdots,n-1\}\cup\{p_1,\cdots,p_m\}$. The double copy structure \cite{Kawai:1985xq,Bern:2008qj,Chiodaroli:2014xia,Johansson:2015oia,Johansson:2019dnu} indicates that the coefficient ${\cal C}(\sigma_1,\sigma_3,\epsilon,k)$ depends on polarization vectors $\epsilon_i$ of external gluons,
momenta $k_i$ of all external particles, permutations $\sigma_3$ and $\sigma_1$, but is independent of the permutation $\sigma_2$\footnote{Originally, the double copy means the GR amplitude can be factorized as ${\cal A}_{G}={\cal A}_Y\times {\cal S}\times{\cal A}_Y$, where the kernel ${\cal S}$ is obtained by inverting propagators. Our assumption that the coefficients depend on only one color ordering is equivalent to the original version, as can be observed from \eref{exp-YM-KK} and \eref{exp-G-KK}.}. Thus, suppose we replace $(1,\sigma_2,n)$ by the more general ordering $\sigma$ among all external legs, without fixing $1$ and $n$ at any position, the expansion in \eref{exp-YS-KK} still holds.

The $n$-point YM amplitude ${\cal A}_{Y}(\sigma)$, where $\sigma$ is the color ordering among $n$ external gluons, can be expanded to BAS KK basis as
\bea
{\cal A}_{Y}(\sigma)=\sum_{\sigma_1}\,\hat{\cal C}(\sigma_1,\epsilon_i,k_i)\,{\cal A}_S(1,\sigma_1,n|\sigma)\,.~~~\label{exp-YM-KK}
\eea
Again, $\sigma$ is the color ordering among all external legs, without fixing any one at particular position. The coefficient $\hat{\cal C}(\sigma_1,\epsilon_i,k_i)$ depends on polarization vectors and momenta of external gluons, as well as the permutation $\sigma_1$, but is independent of the color ordering $\sigma$, as implied by the double copy structure.

It is natural to image another type of coefficients $\hat{\cal C}(\sigma,\epsilon_i,k_i)$, which depend on the color ordering $\sigma$ rather than $\sigma_1$, thus one can generalize the expansions \eref{exp-YS-KK} and \eref{exp-YM-KK} to
\bea
{\cal A}_{EY}(1,\sigma_1,n;p_1,\cdots,p_m)=\sum_{\sigma_3}\,\sum_{\sigma_2}\,{\cal C}(\sigma_1,\sigma_3,\epsilon_i,k_i)\,{\cal A}_{S}(1,\sigma_3,n|1,\sigma_2,n)\,\hat{\cal C}(\sigma_2,\W\epsilon_i,k_i)\,,~~~\label{exp-EY-KK}
\eea
and
\bea
{\cal A}_{G}(1,\cdots,n)=\sum_{\sigma_1}\,\sum_{\sigma_2}\,\hat{\cal C}(\sigma_1,\epsilon_i,k_i)\,{\cal A}_S(1,\sigma_1,n|1,\sigma_2,n)\,\hat{\cal C}(\sigma_2,\W\epsilon_i,k_i)\,.~~~\label{exp-G-KK}
\eea
The external particle $i$ which carries both polarization vectors $\epsilon_i$ and $\W\epsilon_i$ is interpreted as the graviton, whose polarization tensor $\varepsilon_i$ can be decomposed as $\varepsilon_i^{\mu\nu}=\epsilon_i^\mu\W\epsilon_i^\nu$. Thus, \eref{exp-EY-KK} is the expansion of the EYM amplitude ${\cal A}_{EY}(1,\sigma_1,n;p_1,\cdots,p_m)$, which contains $n$ external gluons encoded by $i\in\{1,\cdots,n\}$, and $m$ external gravitons encoded by $p_j$ with $j\in\{1,\cdots,m\}$. The expansion \eref{exp-G-KK} is the expansion of the $n$-point GR amplitude ${\cal A}_{G}(1,\cdots,n)$. Notice that in this paper we focus on gravitons of Einstein gravity. In such case, polarization vectors $\W\epsilon_i$ are the same as $\epsilon_i$. However, we still use notations $\epsilon_i$ and $\W\epsilon_i$ to manifest the double copy structure.

One can sum over $\sigma_2$ in \eref{exp-EY-KK} and \eref{exp-G-KK} via the expanded formulas of YM amplitudes in \eref{exp-YM-KK}, resulted in %
\bea
{\cal A}_{EY}(1,\sigma_1,n;p_1,\cdots,p_m)=\sum_{\sigma_3}\,{\cal C}(\sigma_1,\sigma_3,\epsilon_i,k_i)\,{\cal A}_{YM}(1,\sigma_3,n)\,,~~~\label{exp-EY-KK-YM}
\eea
and
\bea
{\cal A}_{G}(\{1,\cdots,n\})=\sum_{\sigma_1}\,\hat{\cal C}(\sigma_1,\epsilon_i,k_i)\,{\cal A}_{YM}(1,\sigma_1,n)\,.~~~\label{exp-G-KK-YM}
\eea
These are the expansions of EYM and GR amplitudes to YM KK basis, which bears strong similarity with expansions \eref{exp-YS-KK}
and \eref{exp-YM-KK}, respectively.

The explicit formulas of ${\cal C}(\sigma_1,\sigma_3,\epsilon_i,k_i)$ and $\hat{\cal C}(\sigma_1,\epsilon_i,k_i)$ were computed via different methods in \cite{Fu:2017uzt,Teng:2017tbo,Du:2017kpo,Du:2017gnh,Feng:2019tvb}. In this paper, we shall reconstruct them by using the constraints from the soft theorems and the universality of soft factors.

\section{Expanded YMS amplitudes, and soft factors for scalar and gluon}
\label{secYS}

In this section, by imposing soft theorems and the universality of soft factors, we reconstruct the expansions of single trace YMS amplitudes to KK BAS basis, as well as the soft factors for the scalar and gluon. The process is as follows. In subsection.\ref{subsecYS1g}, we derive the expansion of the YMS amplitude with only one external gluon, and generalize the leading soft factor for the BAS scalar to the YMS case. In subsection.\ref{subsecsoftg}, we derive the leading and sub-leading soft factors for the gluon from the expansion of YMS amplitude obtained in subsection.\ref{subsecYS1g}. Then, in subsection.\ref{subsecYS2g}, we use the universal soft factors for the scalar and gluon to determine the expansion
of YMS amplitude with two external gluons. Finally, in subsection.\ref{subsecYSng}, we develop a recursive method, which leads to the expansion of general YMS amplitudes. As can be seen, the constraints from soft theorems and universality play the central role throughout the whole process.

\subsection{Expanded YMS amplitude with one gluon and soft factor for scalar}
\label{subsecYS1g}

We start by considering the single trace YMS amplitude ${\cal A}_{YS}(1,\cdots,n;p|\sigma)$, which contains $n$ external
scalars labeled by $i\in\{1,\cdots,n\}$, and a gluon labeled by $p$. Here $\sigma$ denotes an arbitrary color ordering
among all external particles including the gluon, while another color ordering
$(1,\cdots,n)$ includes only scalars. Based on the discussions in subsection.\ref{subsecexpand}, we know that the amplitude
${\cal A}_{YS}(1,\cdots,n;p|\sigma)$ can be expanded to BAS amplitudes as
\bea
{\cal A}_{YS}(1,\cdots,n;p|\sigma)=\sum_{i=1}^{n-1}\,(\epsilon_p\cdot P_i)\,{\cal A}_{S}(1,\cdots,i,p,i+1,\cdots,n|\sigma)\,.~~~~\label{expand-YS-1g}
\eea
We express the coefficients as $\epsilon_p\cdot P_i$, since the amplitude
${\cal A}_{YS}(1,\cdots,n;p|\sigma)$ is a Lorentz invariant which is linear in the polarization vector $\epsilon_p$. The double copy structure indicates
that $P_i$ are independent of the color ordering $\sigma$. In $d$ dimensional space-time, the
coupling constants of BAS and YM theories have mass dimensions $3-d/2$ and $2-d/2$ respectively, thus the mass dimension of $P_i$
must be $1$. It means $P_i$ are combinations of external momenta. Our aim is to determine the combinatory momenta $P_i$ via the soft theorem.

Let us re-scale the external momentum of the leg $1$ as $k_1\to \tau k_1,$ and expand ${\cal A}_{YS}(1,\cdots,n;p|\sigma)$ in $\tau$.
The leading order term is given by
\bea
{\cal A}^{(0)}_{YS}(1,\cdots,n;p|\sigma)=\sum_{i=1}^{n-1}\,(\epsilon_p\cdot P^{(0)}_i)\,{\cal A}^{(0)}_{S}(1,\cdots,i,p,i+1,\cdots,n|\sigma)\,,~~~~\label{soft-YS-1g-0}
\eea
where $P^{(0)}_i$ are leading order contributions of $P_i$. The leading order terms of ${\cal A}_{S}(1,\cdots,i,p,i+1,\cdots,n|\sigma)$
are determined by the leading soft factor for the BAS scalar in \eref{soft-fac-s-0}, namely,
\bea
{\cal A}^{(0)}_{S}(1,\cdots,i,p,i+1,\cdots,n|\sigma)&=&{1\over \tau}\,\Big({\delta_{12}\over s_{12}}+{\delta_{n1}\over s_{n1}}\Big)\,
{\cal A}_{S}(\not{1},2,\cdots,i,p,i+1,\cdots,n|\sigma\setminus1)\,,~~{\rm for}~i\geq2\,,\nn
{\cal A}^{(0)}_{S}(1,p,2,\cdots,n|\sigma)&=&{1\over \tau}\,\Big({\delta_{1p}\over s_{1p}}+{\delta_{n1}\over s_{n1}}\Big)\,
{\cal A}_{S}(\not{1},p,2,\cdots,n|\sigma\setminus1)\,,~~{\rm for}~i=1\,.~~~~\label{soft-s-0}
\eea
Here $\not{1}$ means removing the particle $1$, and $\sigma\setminus1$ stands for the color ordering obtained from $\sigma$ by removing $1$.
Substituting \eref{soft-s-0} into \eref{soft-YS-1g-0}, we get
\bea
{\cal A}^{(0)}_{YS}(1,\cdots,n;p|\sigma)&=&{1\over \tau}\,\Big({\delta_{1p}\over s_{1p}}+{\delta_{n1}\over s_{n1}}\Big)\,(\epsilon_p\cdot P^{(0)}_1)\,
{\cal A}_{S}(\not{1},p,2,\cdots,n|\sigma\setminus1)\nn
& &+{1\over \tau}\,\Big({\delta_{12}\over s_{12}}+{\delta_{n1}\over s_{n1}}\Big)\,\sum_{j=2}^{n-1}\,(\epsilon_p\cdot P^{(0)}_j)\,
{\cal A}_{S}(\not{1},\cdots,j,p,j+1,\cdots,n|\sigma\setminus1)\,.~~~~\label{soft-YS-1g-0-2}
\eea
Since the color ordering $\sigma$ is general, non of $\delta_{n1}$, $\delta_{12}$ and $\delta_{1p}$ can be fixed to be $0$.

The soft theorem requires the factorization
\bea
{\cal A}^{(0)}_{YS}(1,\cdots,n;p|\sigma)=S^{(0)}_s(1)\,{\cal A}_{YS}(\not{1},\cdots,n;p|\sigma\setminus1)\,,~~~~\label{soft-theo-1}
\eea
where $S^{(0)}_s(1)$ is the universal leading soft factor for the scalar $1$. In subsection.\ref{subsecBAS}, we have derived this factor, which is given in \eref{soft-fac-s-0}. However, as discussed previously, since the operator \eref{soft-fac-s-0} is derived from the pure BAS amplitude, it can not tell us whther the operator $S^{(0)}_s(1)$ acts on external gluons. The universality of soft factor requires that $S^{(0)}_s(1)$ always acts
on external BAS scalars in the manner described by \eref{soft-fac-s-0}, therefore the form of $S^{(0)}_s(1)$ in \eref{soft-theo-1} should be
\bea
S^{(0)}_s(1)={1\over\tau}\,\Big({\delta_{12}\over s_{12}}+{\delta_{n1}\over s_{n1}}\Big)+{\cal S}_{s}(1;p)\,,~~\label{require-universal}
\eea
where ${\cal S}_{s}(1;p)$ stands for the operator acts on external gluon $p$.

To determine ${\cal S}_{s}(1;p)$, we use the KK relation to expand ${\cal A}_{S}(p,2,\cdots,n|\sigma\setminus1)$ in the first line of \eref{soft-YS-1g-0-2} as
\bea
{\cal A}_{S}(p,2,\cdots,n|\sigma\setminus1)=-{\cal A}_{S}(2,\{3,\cdots,n-1\}\shuffle p,n|\sigma\setminus1)\,,
\eea
this manipulation turns \eref{soft-YS-1g-0-2} to
\bea
& &{\cal A}^{(0)}_{YS}(1,\cdots,n;p|\sigma)\nn
&=&
{1\over \tau}\,\sum_{j=2}^{n-1}\,\Big[\big(\epsilon_p\cdot P^{(0)}_j\big)\,{\delta_{12}\over s_{12}}+\big(\epsilon_p\cdot (P^{(0)}_j-P^{(0)}_1)\big)\,{\delta_{n1}\over s_{n1}}-\big(\epsilon_p\cdot P^{(0)}_1\big)\,{\delta_{1p}\over s_{1p}}\Big]\,
{\cal A}_{S}(2,\cdots,j,p,j+1,\cdots,n|\sigma\setminus1)\,.~~~~\label{soft-YS-1g-0-3}
\eea
From \eref{soft-YS-1g-0-3}, we see that the condition \eref{require-universal} can be satisfied if and only if
$P^{(0)}_1=0$, therefore ${\cal S}_{s}(1;p)=0$.
Hence, we get the soft factor
\bea
S^{(0)}_s(1)={1\over\tau}\,\Big({\delta_{12}\over s_{12}}+{\delta_{n1}\over s_{n1}}\Big)\,,~~~\label{soft-fac-s-0-YS}
\eea
which is the generalization of \eref{soft-fac-s-0} to the current case.
Since ${\cal S}_{s}(1;p)=0$, the contribution from external gluon is excluded.
Consequently, $P_1$ is the combination of external momenta which vanishes at the leading order, thus is proportional to $k_1$.
We can take $P_1=k_1$ via an overall re-scaling of the amplitude.

\begin{figure}
  \centering
  \includegraphics[width=8cm]{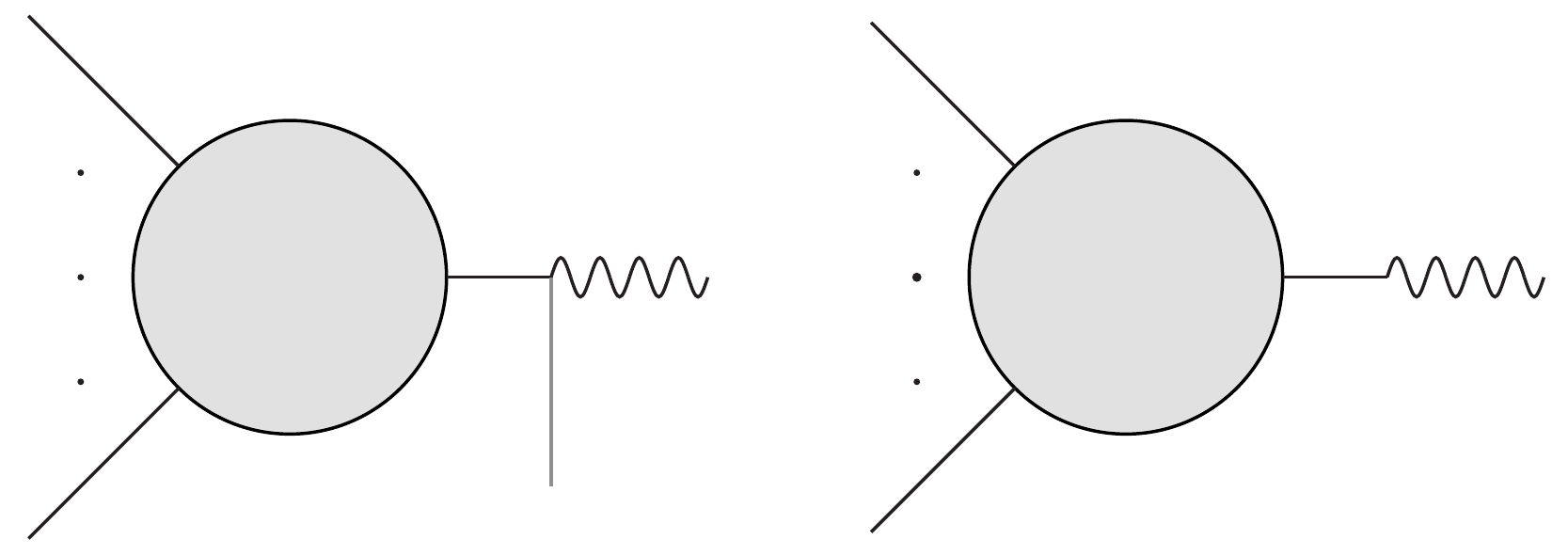} \\
  \caption{Removing the soft scalar from the gluon-scalar-scalar vertex. The gray line stands for the soft scalar.}\label{ssg}
\end{figure}

The phenomenon that the soft factor for the BAS scalar does not act on the gluon can be understood from the Feynman diagram point of view. If one removes the soft scalar from the gluon-scalar-scalar interaction vertex to get the lower-point amplitude, the remaining
diagram means the gluon can be turned to a scalar without any interaction, as can be seen in Figure.\ref{ssg}. This picture is physically unacceptable. Thus, such diagram can never contribute to the soft factor.

The vanishing of $P^{(0)}_1$ indicates that ${\cal A}^{(0)}_{YS}(1,\cdots,n;p|\sigma)$ only comes from the second line at the r.h.s of \eref{soft-YS-1g-0-2}, thus the soft theorem \eref{soft-theo-1} together with the soft factor in \eref{soft-fac-s-0-YS} impose
\bea
& &S^{(0)}_s(1)\,{\cal A}_{YS}(\not{1},\cdots,n;p|\sigma\setminus1)\nn
&=&S^{(0)}_s(1)\,\Big(\sum_{j=2}^{n-1}\,(\epsilon_p\cdot P^{(0)}_j)\,
{\cal A}_{S}(\not{1},\cdots,j,p,j+1,\cdots,n|\sigma\setminus1)\Big)\,,~~~~\label{eq-soft}
\eea
which implies the expansion
\bea
{\cal A}_{YS}(2,\cdots,n;p|\sigma\setminus1)=\sum_{j=2}^{n-1}\,(\epsilon_p\cdot P^{(0)}_j)\,
{\cal A}_{S}(2,\cdots,j,p,j+1,\cdots,n|\sigma\setminus1)\,,~~~~\label{expand-2}
\eea
Comparing the expansion in \eref{expand-2} with that in \eref{expand-YS-1g}, we see that for general $n$
they are totally the same,
up to a relabeling. This observation is based on the condition that the coefficients are independent of the color ordering $\sigma$, which arises from the double copy structure.  Thus, the solution $P_1=k_1$ indicates $P^{(0)}_2=k_2$ in \eref{expand-2}, therefore
\bea
P_2=k_2+\alpha k_1\,.~~~~\label{a}
\eea
To fix the parameter $\alpha$, we consider the soft behavior of the the external scalar $2$.
After taking $k_2\to \tau k_2$ and expanding \eref{expand-YS-1g} in $\tau$, the leading order term is given as
\bea
{\cal A}^{(0)}_{YS}(1,\cdots,n;p|\sigma)&=&{1\over \tau}\,\Big({\delta_{p2}\over s_{p2}}+{\delta_{23}\over s_{23}}\Big)\,(\epsilon_p\cdot P_1^{(0)})\,
{\cal A}_S(1,p,\not{2},\cdots,n|\sigma\setminus2)\nn
& &+{1\over \tau}\,\Big({\delta_{12}\over s_{12}}+{\delta_{2p}\over s_{2p}}\Big)\,(\epsilon_p\cdot P_2^{(0)})\,
{\cal A}_S(1,\not{2},p,\cdots,n|\sigma\setminus2)\nn
& &+{1\over \tau}\,\Big({\delta_{12}\over s_{12}}+{\delta_{23}\over s_{23}}\Big)\,\sum_{i=3}^{n-1}\,(\epsilon_p\cdot P_i^{(0)})\,
{\cal A}_S(1,\not{2},\cdots,i,p,i+1,\cdots,n|\sigma\setminus2)\,.~~~~\label{leading-k2}
\eea
The soft theorem \eref{soft-theo-1} together with the universality of leading order soft factor in \eref{soft-fac-s-0} indicate that
\bea
{\cal A}^{(0)}_{YS}(1,\cdots,n;p|\sigma)={1\over \tau}\,\Big({\delta_{12}\over s_{12}}+{\delta_{23}\over s_{23}}\Big)\,
{\cal A}_{YS}(1,\not{2},\cdots,n;p|\sigma\setminus2)\,.~~\label{e}
\eea
Expanding ${\cal A}_{YS}(1,3,\cdots,n;p|\sigma\setminus2)$ in \eref{e} via the expansion \eref{expand-YS-1g} (with a relabeling), and using the solution $P_1=k_1$, one can observe that
the combination of first two lines at the r.h.s of \eref{leading-k2} gives
\bea
& &{1\over \tau}\,\Big({\delta_{p2}\over s_{p2}}+{\delta_{23}\over s_{23}}\Big)\,(\epsilon_p\cdot P_1^{(0)})\,
{\cal A}_S(1,p,\not{2},\cdots,n|\sigma\setminus2)\nn
& &+{1\over \tau}\,\Big({\delta_{12}\over s_{12}}+{\delta_{2p}\over s_{2p}}\Big)\,(\epsilon_p\cdot P_2^{(0)})
{\cal A}_S(1,\not{2},p,\cdots,n|\sigma\setminus2)\nn
&=&{1\over \tau}\,\Big({\delta_{12}\over s_{12}}+{\delta_{23}\over s_{23}}\Big)\,(\epsilon_p\cdot k_1)\,
{\cal A}_{S}(1,p,3,\cdots,n|\sigma\setminus2)\,,
\eea
which means
\bea
\Big({\delta_{p2}\over s_{p2}}+{\delta_{23}\over s_{23}}\Big)\,(\epsilon_p\cdot P_1^{(0)})
+\Big({\delta_{12}\over s_{12}}+{\delta_{2p}\over s_{2p}}\Big)\,(\epsilon_p\cdot P_2^{(0)})
=\Big({\delta_{12}\over s_{12}}+{\delta_{23}\over s_{23}}\Big)\,(\epsilon_p\cdot k_1)\,.~~~~\label{eq-1}
\eea
For $k_2\to \tau k_2$, we have $P_1^{(0)}=k_1$ and $P_2^{(0)}=\alpha k_1$. Then, we find the solution to \eref{eq-1}
is $\alpha=1$. Here we have used the property $\delta_{ab}=-\delta_{ba}$.

Until now, we have found $P_1=k_1$ and $P_2=k_1+k_2$. Taking the soft limit of other external scalars successively, and applying the same method, we get
\bea
P_i=\sum_{j=1}^i\,k_j\,.~~~~\label{Pi}
\eea
Consequently, the YMS amplitude with one external gluon can be expanded as
\bea
{\cal A}_{YS}(1,\cdots,n;p|\sigma)=(\epsilon_p\cdot X_p)\,{\cal A}_{S}(1,\{2,\cdots,n-1\}\shuffle p,n|\sigma)\,.~~~~\label{expand-YS-1g-2}
\eea
The combinatory momentum $X_p$ is defined as the summation of momenta of legs at the l.h.s of $p$ in the color ordering. The shuffle $\shuffle$ is explained in \eref{shuffle}.

\subsection{Soft factors for gluon}
\label{subsecsoftg}

Form the expansion in \eref{expand-YS-1g-2}, one can determine the leading and sub-leading soft factors for the gluon.
Let us take $k_p\to \tau k_p$, and expand \eref{expand-YS-1g-2} in $\tau$. The leading order term is given as
\bea
{\cal A}^{(0)}_{YS}(1,\cdots,n;p|\sigma)&=&\sum_{i=1}^{n-1}\,{1\over \tau}\,\Big({\delta_{ip}\over s_{ip}}+{\delta_{p(i+1)}\over s_{p(i+1)}}\Big)\,(\epsilon_p\cdot P_i)\,{\cal A}_{S}(1,\cdots,i,\not{p},i+1,\cdots,n|\sigma\setminus p)\nn
&=&{1\over \tau}\,\sum_{i=1}^{n-1}\,\Big[\sum_{j=i}^{n-1}\,\Big({\delta_{jp}\over s_{jp}}+{\delta_{p(j+1)}\over s_{p(j+1)}}\Big)\Big]\,(\epsilon_p\cdot k_i)\,{\cal A}_{S}(1,\cdots,n|\sigma\setminus p)\nn
&=&{1\over \tau}\,\Big[\sum_{i=1}^{n-1}\,\Big({\delta_{ip}\over s_{ip}}+{\delta_{pn}\over s_{pn}}\Big)\,(\epsilon_p\cdot k_i)\Big]\,{\cal A}_{S}(1,\cdots,n|\sigma\setminus p)\nn
&=&{1\over \tau}\,\Big[\sum_{j=1}^n\,{\delta_{jp}\over s_{jp}}\,(\epsilon_p\cdot k_j)\Big]\,{\cal A}_{S}(1,\cdots,n|\sigma\setminus p)\,,~~~~\label{YS-1g-g-0}
\eea
where we have used \eref{Pi} to get the second equality, $\delta_{ab}=-\delta_{ba}$ to get the third one, and the momentum conservation to get the last one. The soft theorem imposes
\bea
{\cal A}^{(0)}_{YS}(1,\cdots,n;p|\sigma)=S^{(0)}_g(p)\,{\cal A}_{S}(1,\cdots,n|\sigma\setminus p)\,.~~~~\label{soft-g-0}
\eea
Comparing \eref{soft-g-0}with the last line at the r.h.s in \eref{YS-1g-g-0}, we find that the leading order soft factor for the gluon is
\bea
S^{(0)}_g(p)={1\over \tau}\,\sum_{j=1}^n\,{\delta_{jp}\,(\epsilon_p\cdot k_j)\over s_{jp}}\,.~~~~\label{soft-fac-g-0}
\eea
There is still an ambiguity that if the operators $S^{(0)}_g(p)$ acts on all external legs, or only on external scalars. This question can not be answered by considering the YMS amplitudes with only one external gluon, and will be solved in the next subsection.

Now we turn to the sub-leading order. The leading order is the $1/\tau$ order, thus the sub-leading order should be $\tau^0$.
To find the $\tau^0$ term ${\cal A}^{(1)}_{YS}(1,\cdots,n;p|\sigma)$, we classify the corresponding Feynman diagrams into two types, and consider them one by one.

The first case, the gluon is coupled to an external scalar $i$ of the pure BAS amplitude ${\cal A}_{S}(1,\cdots,n|\sigma\setminus p)$. Collecting all such diagrams together yields
\bea
B_i(\tau)&=&{\,1\over \tau s_{ip}}\,M_i(\tau)\nn
&=&{\,1\over \tau s_{ip}}\,\big(M_i(0)+\tau\,{\partial\over \partial \tau}\,M_i(\tau)\big|_{\tau=0}+\cdots\big)\,,~~~~\label{B1}
\eea
the first term in the second line describes the leading order soft behavior, therefore
\bea
M_i(0)=\delta_{ip}\,(\epsilon_p\cdot k_i)\,{\cal A}_{S}(1,\cdots,n|\sigma\setminus p)\,.~~~~\label{M}
\eea
The $\tau^0$ contribution of $B_i(\tau)$ arises from the second term in the second line at the r.h.s of \eref{B1}.
In the current case, $\tau$ enters $M_i(\tau)$ only through the combination $k_i+\tau k_p$, this observation indicates
\bea
{\partial\over \partial \tau}\,M_i(\tau)={1\over\tau}\,k_p\cdot{\partial\over \partial k_p}\,M_i(\tau)=k_p\cdot{\partial\over \partial k_i}\,M_i(\tau)\,,
\eea
thus
\bea
{\partial\over \partial \tau}\,M_i(\tau)\big|_{\tau=0}&=&k_p\cdot{\partial\over \partial k_i}\,M_i(0)\nn
&=&\delta_{ip}\,(\epsilon_p\cdot k_i)\,k_p\cdot{\partial\over \partial k_i}\,{\cal A}_{S}(1,\cdots,n|\sigma\setminus p)\,,~~~~\label{M'}
\eea
where we have used \eref{M} to get the second equality. Substituting \eref{M'} into \eref{B1}, and summing over all external scalars $i$, we find the $\tau^0$ term contributed by the first type of Feynman diagrams is given by
\bea
B^{0}=\sum_{i=1}^n\,{\delta_{ip}\,(\epsilon_p\cdot k_i)\over s_{ip}}\,k_p\cdot{\partial\over \partial k_i}\,{\cal A}_{S}(1,\cdots,n|\sigma\setminus p)\,.~~~~\label{B-0}
\eea
\begin{figure}
  \centering
  \includegraphics[width=7cm]{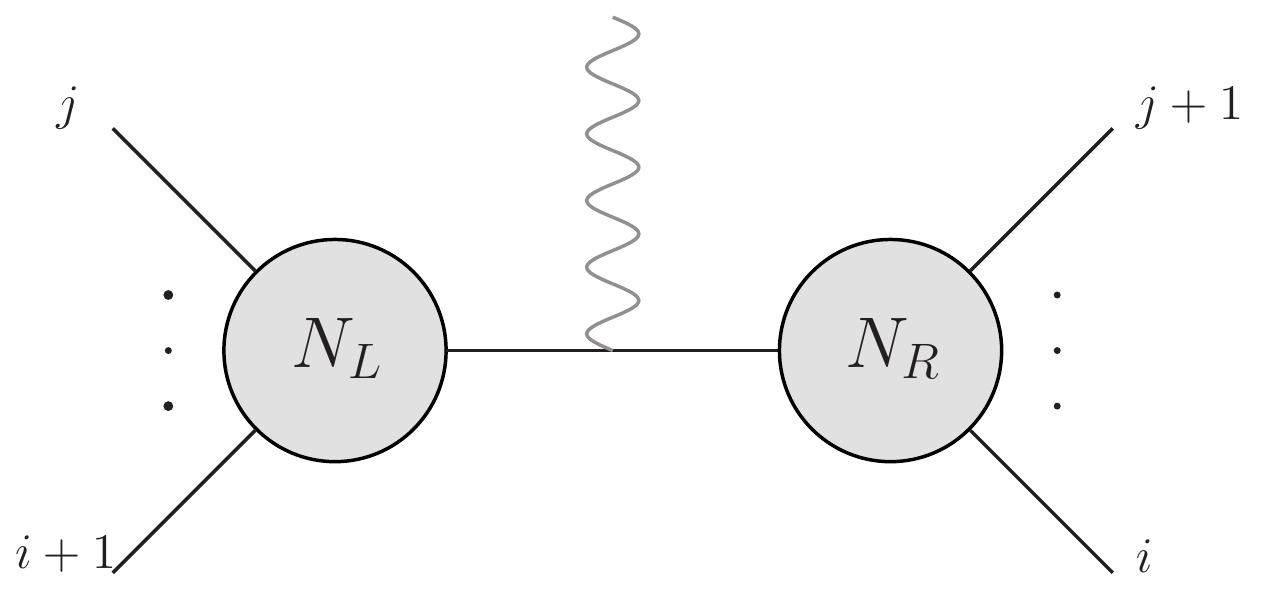} \\
  \caption{The second type of Feynman diagram which contributes to the sub-leading soft operator. The gray wave line denotes the soft gluon.}\label{NsN}
\end{figure}

\begin{figure}
  \centering
  \includegraphics[width=8cm]{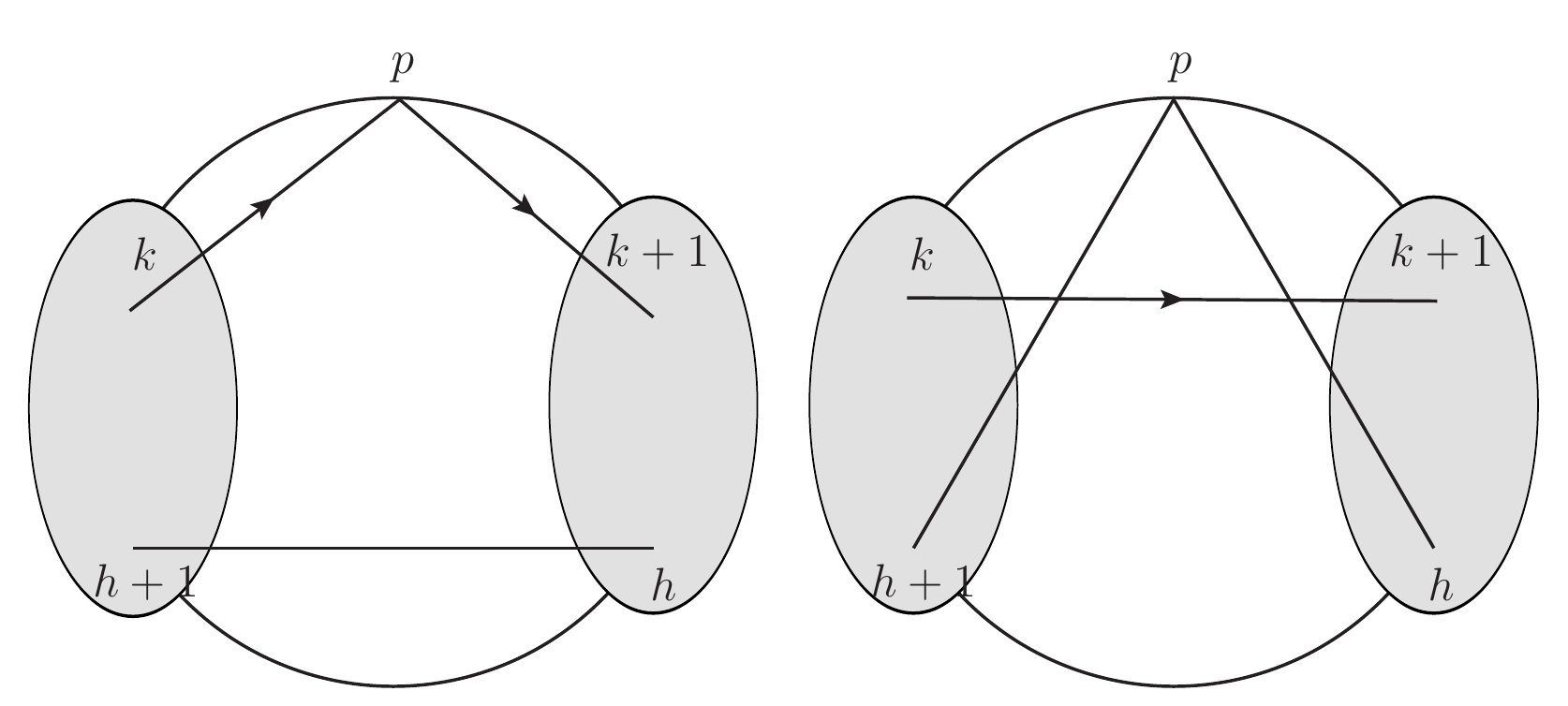} \\
  \caption{Disk diagram for the case $p$ is coupled to $1/s_{(i+1)(i+2)\cdots(j-1)j}$.}\label{2configu}
\end{figure}

The second case, the gluon is coupled to an internal propagator of the BAS amplitude ${\cal A}_{S}(1,\cdots,n|\sigma\setminus p)$, as shown in Figure.\ref{NsN}. In the expansion \eref{expand-YS-1g-2}, each BAS amplitude carries two color orderings $(\{1,\cdots,n\}\shuffle p)$ and $\sigma$. Suppose $p$ is coupled to the propagator $1/s_{(i+1)(i+2)\cdots(j-1)j}$ with $i<j$, only ${\cal A}_{S}(1,\cdots,i,p,i+1,\cdots,n|\sigma)$ and ${\cal A}_{S}(1,\cdots,j,p,j+1,\cdots,n|\sigma)$ in the expansion \eref{expand-YS-1g-2} can carry the correct color orderings. However, this is a necessary condition rather than a sufficient one. As discussed in subsection.\ref{subsecBAS}, if the propagator $1/ s_{(i+1)(i+2)\cdots(j-1)j}$
is contained in the BAS amplitude, then the set of points $\{i+1,\cdots,j\}$ localized on the boundary of disk has only two external lines.
It means the Feynman diagram in Figure.\ref{NsN} requires one of two configurations in Figure.\ref{2configu} to be satisfied. In Figure.\ref{2configu}, $k$
and $h$ can be either $k=i,h=j$ or $k=j,h=i$. The orientations of two disks are the same, and can be either clockwise or anti-clockwise, determined by the color ordering $\sigma$. For general
$\sigma$, ${\cal A}_{S}(1,\cdots,i,p,i+1,\cdots,n|\sigma)$ and ${\cal A}_{S}(1,\cdots,j,p,j+1,\cdots,n|\sigma)$ can ensure neither of two configurations in Figure.\ref{2configu}. This problem will be solved in \eref{D-0}. For now, we just assume that $1/s_{(i+1)(i+2)\cdots(j-1)j}$ is contained in
${\cal A}_{S}(1,\cdots,i,p,i+1,\cdot,n|\sigma)$ and ${\cal A}_{S}(1,\cdots,j,p,j+1,\cdot,n|\sigma)$, and $p$ is coupled to this propagator.

\begin{figure}
  \centering
  \includegraphics[width=10cm]{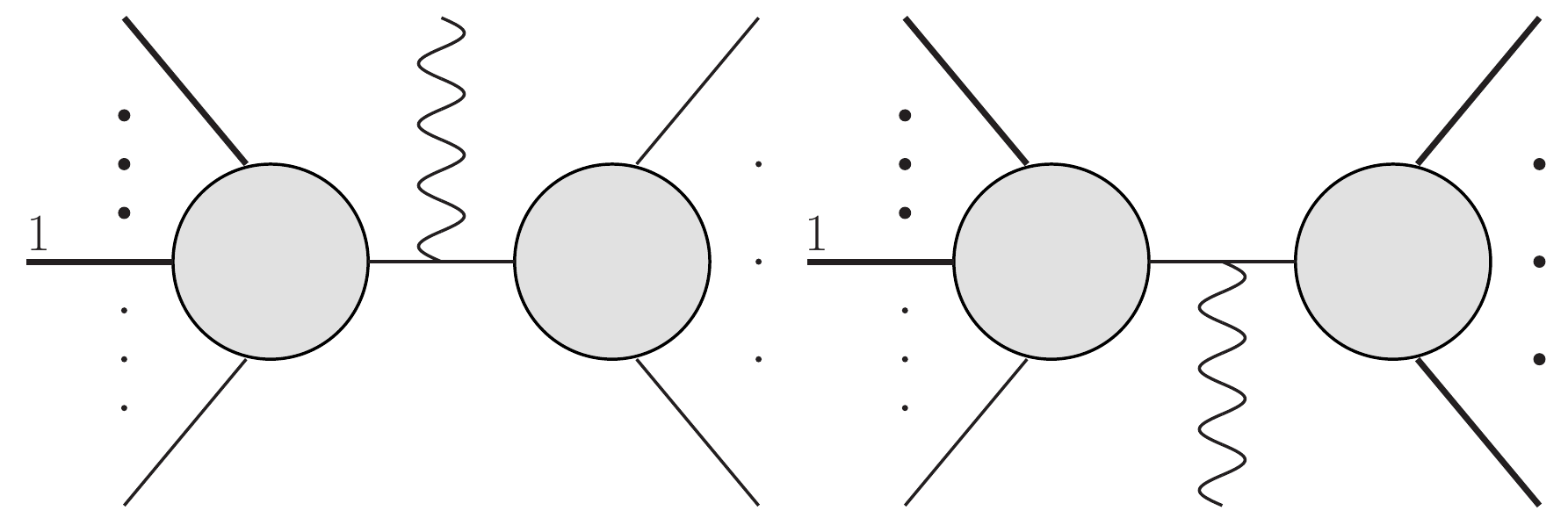} \\
  \caption{Two allowed configurations. The leg $1$ is at left, and the orderings in diagrams are clockwise.}\label{2configu2}
\end{figure}

With the assumption $p$ is coupled to $1/s_{(i+1)(i+2)\cdots(j-1)j}$, let us work out the $\tau^0$ contribution. Collecting contributions from ${\cal A}_{S}(1,\cdots,i,p,i+1,\cdot,n|\sigma)$ and ${\cal A}_{S}(1,\cdots,j,p,j+1,\cdot,n|\sigma)$ gives
\bea
D_{ij}(\tau)&=&{\rm sign}(\pm)\,\Big(\sum_{a=i+1}^j\,\delta_{ap}\Big)\,\big(\epsilon_p\cdot K_{(i+1)j}\big)\,N_L\,{1\over s_{(i+1)(i+2)\cdots(j-1)j}}\,{1\over s_{(i+1)(i+2)\cdots(j-1)jp}}\,N_R(\tau)\,,~~~~\label{NN}
\eea
where
\bea
s_{(i+1)(i+2)\cdots(j-1)jp}=s_{(i+1)(i+2)\cdots(j-1)jp}+2\tau K_{(i+1)j}\cdot k_p\,,
\eea
with $K_{(i+1)j}$ defined in \eref{mandelstam}. Two building blocks $N_L$ and $N_R$ are denoted in Figure.\ref{NsN}. We use $\sum_{a=i+1}^j\,\delta_{ap}$ to ensure that one of legs in $\{i+1,\cdots,j\}$ is adjacent to $p$ in the color ordering $\sigma$, as required by Figure.\ref{2configu}.
The factor $\epsilon_p\cdot K_{(i+1)j}$ arises as follows. Two cases ${\cal A}_{S}(1,\cdots,i,p,i+1,\cdot,n|\sigma)$ and ${\cal A}_{S}(1,\cdots,j,p,j+1,\cdot,n|\sigma)$ correspond to two configurations of Feynman diagrams in Figure.\ref{2configu2}. For either of two configurations,
$X_p$ is the summation of external momenta of bold lines, and we denote $X_p$ for two configurations
as $X_{p;1}$ and $X_{p;2}$, respectively. Since two configurations related to each other by swapping the external line $p$ and propagator
$1/s_{(i+1)(i+2)\cdots(j-1)j}$, the antisymmetry of structure constant $f^{abc}$ indicates a relative $-$ between two cases. Thus we
get $\epsilon_p\cdot (X_{p;1}-X_{p;2})=\pm\epsilon_p\cdot K_{(i+1)j}$, where the combinatory momentum $K_{(i+1)j}$ can be observed from Figure.\ref{2configu2}.
The overall sign denoted by ${\rm sign}(\pm)$ will be treated soon. In \eref{NN}, the $\tau^0$ contribution is just the leading order term obtained by taking $\tau=0$,
thus we get
\bea
D^{0}_{ij}&=&{\rm sign}(\pm)\,\Big(\sum_{a=i+1}^j\,\delta_{ap}\Big)\,{\epsilon_p\cdot K_{(i+1)j}\over s^2_{(i+1)(i+2)\cdots(j-1)j}}\,N_L\,N_R(0)\nn
&=&-{{\rm sign}(\pm)\over2}\,\Big(\sum_{a=i+1}^j\,\delta_{ap}\,\epsilon_p\cdot {\partial\over\partial k_a}\,{1\over s_{(b+1)(b+2)\cdots(a-1)a}}\Big)\,N_L\,N_R(0).~~~~\label{D0}
\eea
Here a subtle point is that the Mandelstam variable $s_{(i+1)(i+2)\cdots(j-1)j}$ defined in \eref{mandelstam} contains $k_a^2$. Although $k_a^2$ vanish due to the on-shell condition, they contribute $2k_a^\mu$ when taking the derivative of $k_{a\mu}$.
Summing over $D_{ij}$ provides the $\tau^0$
contribution for the second case,
\bea
D^0&=&\sum_{i\in\{1,\cdots,n\}}\,\sum_{j\in\{1,\cdots,n\}\setminus i}\,D^{0}_{ij}\nn
&=&-{1\over 2}\,\sum_{a=1}^n\,\delta_{ap}\,\epsilon_p\cdot {\partial\over\partial k_a}\,{\cal A}_{S}(1,\cdots,n|\sigma\setminus p)\nn
&=&-\sum_{a=1}^n\,{\delta_{ap}\,(k_p\cdot k_a)\over s_{ap}}\,\epsilon_p\cdot {\partial\over\partial k_a}\,{\cal A}_{S}(1,\cdots,n|\sigma\setminus p)\,.~~~~\label{D-0}
\eea
The equivalence between $s_{(i+1)(i+2)\cdots(j-1)j}$ and $s_{(j+1)(j+2)\cdots(i-1)i}$ turns the summation of $a$ from $\sum_{a=i+1}^j$ to $\sum_{a=1}^n$. The overall sign ${\rm sign}(\pm)$ in \eref{NN} and \eref{D0} is absorbed by ${\cal A}_{S}(1,\cdots,n|\sigma\setminus p)$, based on our convention that removing the external leg $p$ from ${\cal A}_{S}(1,\cdots,k,p,k+1,\cdots,n|\sigma)$ generates ${\cal A}_{S}(1,\cdots,k,\not{p},k+1,\cdots,n|\sigma\setminus p)$, without any relative sign, as interpreted in subsection.\ref{subsecBAS}. When taking the derivative, one need not to worry about if one of two configurations in Figure.\ref{2configu} is satisfied, since the un-allowed propagators will not contribute.

The soft theorem requires
\bea
{\cal A}^{(1)}_{YS}(1,\cdots,n;p|\sigma)=S^{(1)}_g(p)\,{\cal A}_{S}(1,\cdots,n|\sigma\setminus p)\,.
\eea
Combining \eref{B-0} and \eref{D-0} together, we find
\bea
S^{(1)}_g(p)=-\sum_{i=1}^n\,{\delta_{ip}\over s_{ip}}\,k_i\cdot f_p\cdot {\partial\over\partial k_i}\,,~~~~\label{soft-fac-g-1}
\eea
where $f_a^{\mu\nu}$ is the field strength tensor defined as $f_a^{\mu\nu}\equiv k^\mu_a\epsilon^\nu_a-\epsilon^\mu_a k^\nu_a$.
At this step, there are two ambiguities. First, if the sub-leading soft operator $S^{(1)}_g(p)$ acts on all external legs, or only on external scalars. Secondly, suppose $S^{(1)}_g(p)$ also acts on external gluons, it is not clear that if it acts only on external momenta, or on all Lorentz
vectors including polarization vectors. Such ambiguities will be clarified in the next subsection, by considering the YMS amplitude with two external gluons.

\subsection{Expanded YMS amplitudes with two gluons}
\label{subsecYS2g}

In the previous two subsections, we have figured out the expansion of YMS amplitudes with one external gluon to BAS amplitudes. The leading soft factor for the scalar, the leading and sub-leading soft factors for the gluon, are also obtained. In this subsection, we show that by imposing the soft theorem, and the universality of soft factors in \eref{soft-fac-s-0}, \eref{soft-fac-g-0} and \eref{soft-fac-g-1}, one can find the expansion of the YMS amplitudes with two external gluons. The soft factors for the gluon provided in \eref{soft-fac-g-0} and \eref{soft-fac-g-1} have some ambiguities, as pointed out in the previous subsection. These ambiguities will also be solved in this subsection.

Consider the expansion of the YMS amplitude ${\cal A}_{YS}(1,\cdots,n;p,q|\sigma)$,
which contains $n$ external scalars encoded by $i\in\{1,\cdots,n\}$, and two gluons labeled by $p$ and $q$. The Lorentz invariance, the linearity in $\epsilon_p$ and $\epsilon_q$, together with the counting of mass dimension, indicate that ${\cal A}_{YS}(1,\cdots,n;p,q|\sigma)$
can be expanded as follows,
\bea
{\cal A}_{YS}(1,\cdots,n;p,q|\sigma)&=&(\epsilon_p\cdot {\cal X}_p)\,(\epsilon_q\cdot{\cal X}_q)\,{\cal A}_S(1,\{2,\cdots,n-1\}\shuffle p\shuffle q,n|\sigma)\nn
& &+(\epsilon_p\cdot\epsilon_q)\,{\cal Z}_{pq}\,{\cal A}_S(1,\{2,\cdots,n-1\}\shuffle p\shuffle q,n|\sigma)\,,~~~~\label{expand-YS-2g}
\eea
where ${\cal X}_p$ and ${\cal X}_q$ are the combinations of external momenta, while ${\cal Z}_{pq}$ are the combinations of the contractions of external momenta. We first use the soft factor for the scalar to fix ${\cal X}_p$ and ${\cal X}_q$, the method is similar to that used in subsection.\ref{subsecYS1g}. Taking $k_1\to \tau k_1$ and expanding in $\tau$ gives
\bea
{\cal A}^{(0)}_{YS}(1,\cdots,n;p,q|\sigma)&=&{1\over \tau}\,\Big({\delta_{1p}\over s_{1p}}+{\delta_{n1}\over s_{n1}}\Big)(\epsilon_p\cdot {\cal X}^{(0)}_p)\,(\epsilon_q\cdot{\cal X}^{(0)}_q)\,\,{\cal A}_S(\not{1},p,\{2,\cdots,n-1\}\shuffle q,n|\sigma\setminus1)\nn
& &+{1\over \tau}\,\Big({\delta_{1q}\over s_{1q}}+{\delta_{n1}\over s_{n1}}\Big)\,(\epsilon_p\cdot {\cal X}^{(0)}_p)\,(\epsilon_q\cdot{\cal X}^{(0)}_q)\,{\cal A}_S(\not{1},q,\{2,\cdots,n-1\}\shuffle p,n|\sigma\setminus1)\nn
& &+{1\over \tau}\,\Big({\delta_{12}\over s_{12}}+{\delta_{n1}\over s_{n1}}\Big)\,(\epsilon_p\cdot {\cal X}^{(0)}_p)\,(\epsilon_q\cdot{\cal X}^{(0)}_q)\,{\cal A}_S(\not{1},2,\{3,\cdots,n-1\}\shuffle p\shuffle q,n|\sigma\setminus1)\nn
& &+\epsilon_p\cdot\epsilon_q\,{\rm terms}\,.~~~~\label{expand-YS-2g-2}
\eea
The soft theorem imposes
\bea
{\cal A}^{(0)}_{YS}(1,\cdots,n;p,q|\sigma)&=&S^{(0)}_s(1)\,{\cal A}^{(0)}_{YS}(\not{1},\cdots,n;p,q|\sigma\setminus1)\nn
&=&{1\over \tau}\,\Big({\delta_{12}\over s_{12}}+{\delta_{n1}\over s_{n1}}\Big)\,{\cal A}^{(0)}_{YS}(\not{1},\cdots,n;p,q|\sigma\setminus1)\,,~~~\label{poles1i}
\eea
where we have used the universality of $S^{(0)}_s(1)$ in \eref{soft-fac-s-0}. From the second line at the r.h.s of \eref{poles1i} we see that
poles $s_{1p}$ and $s_{1q}$ can not enter ${\cal A}^{(0)}_{YS}(1,\cdots,n;p,q|\sigma)$, thus the first and second lines at the r.h.s of \eref{expand-YS-2g-2} must vanish. Thus, for the color orderings $(1,p,\cdots,n)$, the combinatory momentum ${\cal X}_p$ is given as
\bea
{\cal X}_p\big|_{1,p,\cdots,n}=k_1\,.~~~~\label{Xp1}
\eea
Similarly, for the color orderings $(1,q,\cdots,n)$, the combinatory momentum ${\cal X}_q$ is
\bea
{\cal X}_q\big|_{1,q,\cdots,n}=k_1\,.~~~~\label{Xq1}
\eea
Solutions \eref{Xp1} and \eref{Xq1} fixes ${\cal X}^{(0)}_p\big|_{\not{1},2,p,\cdots,n}$ and ${\cal X}^{(0)}_q\big|_{\not{1},2,p,\cdots,n}$
to be
\bea
{\cal X}^{(0)}_p\big|_{\not{1},2,p,\cdots,n}=k_2\,,~~~~{\cal X}^{(0)}_q\big|_{\not{1},2,p,\cdots,n}=k_2\,,
\eea
which indicates
\bea
{\cal X}_p\big|_{1,2,p,\cdots,n}=k_2+\alpha_1k_1\,,~~~~{\cal X}_q\big|_{1,2,p,\cdots,n}=k_2+\alpha_2k_2\,,
\eea
the reason is the same as that for obtaining \eref{a} in subsection.\ref{subsecYS1g}.
Similar as in subsection.\ref{subsecYS1g}, one can consider the leading order soft behavior of the external scalar $2$ to fix $\alpha_1$
and $\alpha_2$ as
$\alpha_1=\alpha_2=1$. Taking the soft limit of external scalars successively, and repeating the same procedure, we arrive at
\bea
{\cal X}_p\big|_{1,\cdots,i,p,i+1,\cdots,q,\cdots,n}=\sum_{a=1}^{i}\,k_a\,,~~~~{\cal X}_q\big|_{1,\cdots,j,q,j+1,\cdots,p,\cdots,n}=\sum_{b=1}^{j}\,k_b\,.
\eea
We still need to workout ${\cal X}_q\big|_{1,\cdots,p,\cdots,q,\cdots,n}$ and ${\cal X}_p\big|_{1,\cdots,q,\cdots,p,\cdots,n}$.
This goal can be achieved by considering the soft behavior of external scalars $n,n-1,n-2,\cdots$ successively and applying the same method, resulted in
\bea
& &{\cal X}_q\big|_{1,\cdots,p,\cdots,i,q,i+1,\cdots,n}=-\sum_{a=i+1}^{n}\,k_a=k_p+k_q+\sum_{a'=1}^{i}\,k_{a'}\doteq k_p+\sum_{a'=1}^{i}\,k_{a'}\,,\nn
& &{\cal X}_p\big|_{1,\cdots,q,\cdots,j,p,j+1,\cdots,n}=-\sum_{b=j+1}^{n}\,k_b=k_p+k_q+\sum_{b'=1}^{j}\,k_{b'}\doteq k_q+\sum_{b'=1}^{j}\,k_{b'}\,.
\eea
Here $\doteq$ in the first line means objects at two sides are equivalent to each other when contracting with $k_q$, while $\doteq$
in the second line means the equivalence when contracting with $k_p$.
Consequently, we find
\bea
{\cal X}_p=X_p\,,~~~~~~~~{\cal X}_q=X_q\,,~~~~\label{XpXq}
\eea
where the combinatory momentum $X_p$ is defined at the end of subsection.\ref{subsecYS1g}.

The first line at the r.h.s of \eref{expand-YS-2g} has been fixed by the solution \eref{XpXq}, now we need to determine ${\cal Z}_{pq}$ in the second line.
To do so, we consider the sub-leading order soft behavior of the gluon $q$, by employing the sub-leading soft operator \eref{soft-fac-g-1}. As mentioned in the previous subsection, the sub-leading soft operator \eref{soft-fac-g-1} has some ambiguities. Fortunately, such ambiguities can be solved by the solution \eref{XpXq}. To start, we regroup the first line at the r.h.s of \eref{expand-YS-2g} as
\bea
& &(\epsilon_p\cdot X_p)\,(\epsilon_q\cdot X_q)\,{\cal A}_S(1,\{2,\cdots,n-1\}\shuffle p\shuffle q,n|\sigma)\nn
&=&(\epsilon_p\cdot Y_p)\,(\epsilon_q\cdot X_q)\,{\cal A}_S(1,\{2,\cdots,n-1\}\shuffle p\shuffle q,n|\sigma)\nn
& &+(\epsilon_p\cdot k_q)\,(\epsilon_q\cdot X_q)\,{\cal A}_S(1,\{2,\cdots,n-1\}\shuffle \{q,p\},n|\sigma)\nn
&=&(\epsilon_p\cdot Y_p)\,{\cal A}_{YS}(1,\{2,\cdots,n-1\}\shuffle q,n;p|\sigma)\nn
& &+(\epsilon_p\cdot k_q)\,(\epsilon_q\cdot X_q)\,{\cal A}_S(1,\{2,\cdots,n-1\}\shuffle \{q,p\},n|\sigma)\,,
\eea
since the Lorentz invariant $\epsilon_p\cdot k_q$ contributes $\tau$ when considering the soft behavior of the gluon $q$.
The combinatory momentum $Y_p$ is defined as the summation of momenta of only scalars at the l.h.s of $p$ in the color ordering.
Then, we take $k_q\to \tau k_q$ and expand \eref{expand-YS-2g} in $\tau$ to get
\bea
{\cal A}^{(1)}_{YS}(1,\cdots,n;p,q|\sigma)&=&(\epsilon_p\cdot Y_p)\,{\cal A}^{(1)}_{YS}(1,\{2,\cdots,n-1\}\shuffle p,n;q|\sigma)\nn
& &+\tau\,(\epsilon_p\cdot k_q)\,(\epsilon_q\cdot X_q)\,{\cal A}^{(0)}_S(1,\{2,\cdots,n-1\}\shuffle \{q,p\},n|\sigma)\nn
& &+\epsilon_p\cdot\epsilon_q\,{\rm terms}\,.~~~~\label{A1-1}
\eea

The soft theorem requires
\bea
{\cal A}^{(1)}_{YS}(1,\cdots,n;p,q|\sigma)&=&S^{(1)}_g(q)\,{\cal A}_{YS}(1,\cdots,n;p|\sigma\setminus q)\nn
&=&S^{(1)}_g(q)\,\big[(\epsilon_p\cdot X_p)\,{\cal A}_S(1,\{2,\cdots,n-1\}\shuffle p,n|\sigma\setminus q)\big]\nn
&=&\big[S^{(1)}_g(q)\,(\epsilon_p\cdot X_p)\big]\,{\cal A}_S(1,\{2,\cdots,n-1\}\shuffle p,n|\sigma\setminus q)\nn
& &+(\epsilon_p\cdot X_p)\,\big[S^{(1)}_g(q)\,{\cal A}_S(1,\{2,\cdots,n-1\}\shuffle p,n|\sigma\setminus q)\big]\,,~~~~\label{A1-2}
\eea
where we have used the expansion \eref{expand-YS-1g-2} to get the second equality, and the Leibnitz's rule to get the third.
Comparing \eref{A1-2} with \eref{A1-1} gives
\bea
& &\big[S^{(1)}_g(q)\,(\epsilon_p\cdot X_p)\big]\,{\cal A}_S(1,\{2,\cdots,n-1\}\shuffle p,n|\sigma\setminus q)\nn
&=&\tau\,(\epsilon_p\cdot k_q)\,(\epsilon_q\cdot X_q)\,{\cal A}^{(0)}_S(1,\{2,\cdots,n-1\}\shuffle \{q,p\},n|\sigma)
+\epsilon_p\cdot\epsilon_q\,{\rm terms}\,,~~~~\label{equ-soft}
\eea
since the soft theorem ensures
\bea
{\cal A}^{(1)}_{YS}(1,\{2,\cdots,n-1\}\shuffle p,n;q|\sigma)&=&S^{(1)}_g(q)\,{\cal A}_{YS}(1,\{2,\cdots,n-1\}\shuffle p,n|\sigma\setminus q)\,,
\eea
and $Y_p=X_p$ for pure BAS amplitudes. Thus one can figure out the second and third lines at the r.h.s of \eref{A1-1}
by computing $S^{(1)}_g(q)\,(\epsilon_p\cdot X_p)$. Using the formula of $S^{(1)}_g(q)$ in \eref{soft-fac-g-1}, it is straightforward to get
\bea
\epsilon_p\cdot\big(S^{(1)}_g(q)\,X_p\big)&=&\sum_{i=1}^j\,{\delta_{iq}\over s_{iq}}\,(\epsilon_p\cdot f_q\cdot k_i)\,,~~~~{\rm for}~X_p=\sum_{i=1}^j\,k_i\,.
\eea
The above result does not contain the pole $s_{pq}$, which is manifestly included in ${\cal A}^{(0)}_S(1,\{2,\cdots,n-1\}\shuffle \{q,p\},n|\sigma)$ in \eref{equ-soft}. This problem can be solved only if $S^{(1)}_g(q)$ also acts on the polarization vector $\epsilon_p$.
In other words, the correct universal formula of $S^{(1)}_g(q)$ is
\bea
S^{(1)}_g(q)&=&-\sum_{V_a}\,{\delta_{aq}\over s_{aq}}\,V_a\cdot f_q\cdot{\partial\over\partial V_a}\nn
&=&\sum_a\,{\delta_{aq}\over s_{aq}}\,(\epsilon_q\cdot J_a\cdot k_q)\,,~~~~\label{soft-fac-g-1-1}
\eea
where $a$ denotes an external leg which can be either a scalar or a gluon, and $V_a$ denotes the Lorentz vector carried by the leg $a$
which can be either a momentum or a polarization vector. In the first line at the r.h.s, the summation is over all $V_a$. In the second line,
the summation is over all external legs $a$, and $J_{a}^{\mu\nu}$ stands for the angular momentum operator for the leg $a$\footnote{The angular momentum operator $J_a^{\mu\nu}$ acts on Lorentz vector $k^\rho_a$ with the orbital part of the generator and on $\epsilon^\rho_a$ with the spin part of the generator in the vector representation as follows,
\bea
J_a^{\mu\nu}\,k_a^\rho=k_a^{[\mu}\,{\partial k_a^\rho\over\partial k_{a,\nu]}}\,,~~~~
J_a^{\mu\nu}\,\epsilon_a^\rho=\big(\eta^{\nu\rho}\,\delta^\mu_\sigma-\eta^{\mu\rho}\,\delta^\nu_\sigma\big)\,\epsilon^\sigma_a\,.
\eea
These actions can be summarized as in the first line at the r.h.s of \eref{soft-fac-g-1-1}, due to the observation that the amplitude is linear in each polarization vector.}. Now the ambiguity for the formula \eref{soft-fac-g-1} has been clarified. Notice that the universality of the soft factor $S^{(1)}_g(q)$ has been used implicitly when generating \eref{soft-fac-g-1} to \eref{soft-fac-g-1-1}.

Using the soft operator in \eref{soft-fac-g-1-1}, we immediately get
\bea
S^{(1)}_g(q)\,(\epsilon_p\cdot X_p)&=&-{\delta_{pq}\over s_{pq}}\,(\epsilon_p\cdot f_q\cdot X_p)+\sum_{i=1}^j\,{\delta_{iq}\over s_{iq}}\,(\epsilon_p\cdot f_q\cdot k_i)\nn
&=&\sum_{i=1}^j\,\Big({\delta_{iq}\over s_{iq}}-{\delta_{pq}\over s_{pq}}\Big)\,(\epsilon_p\cdot f_q\cdot k_i)\nn
&=&\sum_{i=1}^j\,\Big[\sum_{l=i}^{j-1}\,\Big({\delta_{lq}\over s_{lq}}+{\delta_{q(l+1)}\over s_{q(l+1)}}\Big)+{\delta_{jq}\over s_{jq}}+{\delta_{qp}\over s_{pq}}\Big]\,(\epsilon_p\cdot f_q\cdot k_i)\nn
&=&\sum_{l=1}^{j}\,\Big({\delta_{lq}\over s_{lq}}+{\delta_{q(l+1)}\over s_{q(l+1)}}\Big)\,\Big[\epsilon_p\cdot f_q\cdot\Big(\sum_{i=1}^l\,k_i\Big)\Big],~~~~\label{SeX}
\eea
where $\delta_{ab}=-\delta_{ba}$ has been used to get the third equality. In the last line, $\delta_{q(j+1)}$ and $s_{q(j+1)}$
should be understood as $j+1=p$. The reason for organizing $S^{(1)}_g(q)\,(\epsilon_p\cdot X_p)$ in the above way is as follows.
Both $(\epsilon_p\cdot f_q\cdot X_p)$ and $(\epsilon_p\cdot f_q\cdot k_i)$ carry $\tau$ automatically when taking $k_q\to \tau k_q$, but ${\cal A}^{(1)}_{YS}(1,\cdots,n;p,q|\sigma)$ is at the $\tau^0$ order, thus the BAS amplitudes in the expansion provide the leading order contributions
to cancel $\tau$. In order to extract such leading order contributions, we rewrite $S^{(1)}_g(q)\,(\epsilon_p\cdot X_p)$ to manifest the factors
\bea
{\delta_{lq}\over s_{lq}}+{\delta_{q(l+1)}\over s_{q(l+1)}}\,,
\eea
which are proportional to the leading order soft factors for scalars given in \eref{soft-fac-s-0}.
With $S^{(1)}_g(q)\,(\epsilon_p\cdot X_p)$ expressed in \eref{SeX}, we have
\bea
& &\Big[S^{(1)}_g(q)\,(\epsilon_p\cdot X_p)\Big]\,{\cal A}_S(1,\{2,\cdots,n-1\}\shuffle p,n|\sigma\setminus q)\nn
&=&\sum_{j=1}^{n-1}\,\sum_{l=1}^{j}\,\Big({\delta_{lq}\over s_{lq}}+{\delta_{q(l+1)}\over s_{q(l+1)}}\Big)\,\Big[\epsilon_p\cdot f_q\cdot\Big(\sum_{i=1}^l\,k_i\Big)\Big]\,{\cal A}_S(1,\cdots,j,p,j+1,\cdots,n|\sigma\setminus q)\nn
&=&\sum_{j=1}^{n-1}\,\sum_{l=1}^{j}\,\tau\,\Big[\epsilon_p\cdot f_q\cdot\Big(\sum_{i=1}^l\,k_i\Big)\Big]\,{\cal A}^{(0)}_S(1,\cdots,l,q,l+1,\cdots,j,p,j+1,\cdots,n|\sigma)\nn
&=&\tau\,(\epsilon_p\cdot f_q\cdot X_q)\,{\cal A}^{(0)}_S(1,\{2,\cdots,n-1\}\shuffle\{q,p\},n|\sigma)\,.~~~~\label{SeXA}
\eea
The second equality is obtained by employing the soft theorem
\bea
& &{\cal A}^{(0)}_S(1,\cdots,l,q,l+1,\cdots,j,p,j+1,\cdots,n|\sigma)\nn
&=&{1\over \tau}\,\Big({\delta_{lq}\over s_{lq}}+{\delta_{q(l+1)}\over s_{q(l+1)}}\Big)\,{\cal A}_S(1,\cdots,j,p,j+1,\cdots,n|\sigma\setminus q)\,,
\eea
with the universal soft factor for the scalar in \eref{soft-fac-s-0}. The third one is obtained via the definition of $X_q$ and $\shuffle$.
Now the unknown $\epsilon_p\cdot\epsilon_q$ terms in and \eref{equ-soft} and \eref{A1-1} have been fixed by \eref{SeXA}.
Substituting \eref{SeXA} and \eref{equ-soft} into \eref{A1-1}, we obtain the sub-leading order soft behavior
\bea
{\cal A}^{(1)}_{YS}(1,\cdots,n;p,q|\sigma)&=&(\epsilon_p\cdot Y_p)\,{\cal A}^{(1)}_{YS}(1,\{2,\cdots,n-1\}\shuffle p,n;q|\sigma)\nn
& &+\tau\,(\epsilon_p\cdot f_q\cdot X_q)\,{\cal A}^{(0)}_S(1,\{2,\cdots,n-1\}\shuffle \{q,p\},n|\sigma)\,,~~~~\label{soft-A1}
\eea
which indicates the expansion
\bea
{\cal A}_{YS}(1,\cdots,n;p,q|\sigma)&=&(\epsilon_p\cdot Y_p)\,{\cal A}_{YS}(1,\{2,\cdots,n-1\}\shuffle p,n;q|\sigma)\nn
& &+(\epsilon_p\cdot f_q\cdot X_q)\,{\cal A}_S(1,\{2,\cdots,n-1\}\shuffle \{q,p\},n|\sigma)\,.~~~~\label{expand-recur-YS-2g}
\eea
This is the recursive expansion found in \cite{Fu:2017uzt}\footnote{In \cite{Fu:2017uzt}, the recursive expansion is for the single trace EYM amplitudes. As will be seen in section.\ref{secEYMGR}, the recursive expansion for EYM amplitudes is extremely similar as that for YMS ones.}. One can get the expansion of ${\cal A}_{YS}(1,\cdots,n;p,q|\sigma)$ to BAS basis by expanding ${\cal A}_{YS}(1,\{2,\cdots,n-1\}\shuffle p,n;q|\sigma)$ via \eref{expand-YS-1g-2}.

One may wonder if ${\cal Z}_{pq}$ in \eref{expand-YS-2g} contain terms at $\tau^2$ order when taking $k_q\to \tau k_q$, these terms can not be detected by considering the sub-leading order soft contribution of the gluon $q$. Such possibility can be excluded via the following argument.
The mass dimension of each ${\cal Z}_{pq}$ is $2$, thus ${\cal Z}_{pq}$ should be the combination of the contractions of external momenta. Since the on-shell condition imposes $k_q^2=0$, ${\cal Z}_{pq}$ can not include the $\tau^2$ term.

As can be seen in \cite{Fu:2017uzt}, with the solution \eref{XpXq} on hand, the second line at the r.h.s of \eref{expand-recur-YS-2g} can be fixed by imposing the gauge invariance condition to the gluon $q$. In our method, this condition has not been used. Indeed, the gauge invariance of $q$ is ensured by the gauge invariance of the soft operator \eref{soft-fac-g-1-1}. When replacing $\epsilon_q$ by $k_q$, the operator \eref{soft-fac-g-1-1} vanishes due to the antisymmetry of $J_a^{\mu\nu}$.

Before ending this subsection, let us solve the ambiguity of the leading order soft factor $S^{(0)}_g(a)$ in \eref{soft-fac-g-0}. From the expansion \eref{expand-recur-YS-2g}, one can find the leading order soft behavior of ${\cal A}_{YS}(1,\cdots,n;p,q|\sigma)$ as
\bea
& &{\cal A}^{(0)}_{YS}(1,\cdots,n;p,q|\sigma)\nn
&=&{1\over \tau}\,\sum_{i=1}^{n-1}\,\Big[\sum_{j\in\{1,\cdots,n-1\}\setminus i}\,\Big({\delta_{jq}\over s_{jq}}+{\delta_{q(j+1)}\over s_{q(j+1)}}\Big)\,(\epsilon_p\cdot Y_p)\,(\epsilon_q\cdot X_q)\,{\cal A}_S(1,\cdots,j,\not{q},j+1,\cdots,i,p,i+1,\cdots,n|\sigma)\nn
& &+\Big({\delta_{iq}\over s_{iq}}+{\delta_{qp}\over s_{qp}}\Big)\,(\epsilon_p\cdot Y_p)\,(\epsilon_q\cdot X_q)\,{\cal A}_S(1,\cdots,i,\not{q},p,i+1,\cdots,n|\sigma)\nn
& &+\Big({\delta_{pq}\over s_{pq}}+{\delta_{q(i+1)}\over s_{q(i+1)}}\Big)\,(\epsilon_p\cdot Y_p)\,(\epsilon_q\cdot X_q)\,{\cal A}_S(1,\cdots,i,p,\not{q},i+1,\cdots,n|\sigma)\Big]\nn
&=&{1\over \tau}\,\Big[{\delta_{pq}\,(\epsilon_q\cdot k_p)\over s_{pq}}+{\delta_{qn}\,(\epsilon_q\cdot k_p)\over s_{qn}}
+\sum_{j=1}^{n-1}\,\Big({\delta_{jq}\,(\epsilon_q\cdot k_j)\over s_{jq}}+{\delta_{qn}\,(\epsilon_q\cdot k_j)\over s_{qn}}\Big)\Big]\,
{\cal A}_{YS}(1,\cdots,n;p|\sigma)\nn
&=&{1\over \tau}\,\Big({\delta_{pq}\,(\epsilon_q\cdot k_p)\over s_{pq}}+\sum_{j=1}^n\,{\delta_{jq}\,(\epsilon_q\cdot k_j)\over s_{jq}}\Big)\,{\cal A}_{YS}(1,\cdots,n;p|\sigma)\,.~~~~\label{1order-YS2g-g}
\eea
The first equality is obtained by expanding ${\cal A}_{YS}(1,\{2,\cdots,n-1\}\shuffle p,n;q|\sigma)$ and applying the soft theorem to get the
leading order contributions of BAS amplitudes. The second arises from $\delta_{ab}=-\delta_{ba}$, and the expansion of ${\cal A}_{YS}(1,\cdots,n;p)$ in \eref{expand-YS-1g-2}. The last one uses momentum conservation. From \eref{1order-YS2g-g}, one can extract the soft factor $S^{(0)}_g(q)$ as
\bea
S^{(0)}_g(q)={1\over \tau}\,\sum_{a\in\sigma}\,{\delta_{aq}\,(\epsilon_q\cdot k_a)\over s_{aq}}\,,~~~~\label{soft-fac-g-0-1}
\eea
the summation is over all external legs $a$ included in the color ordering $\sigma$, thus the ambiguity has been clarified. Notice that the operator \eref{soft-fac-g-0-1} is gauge invariant. After taking $\epsilon_p\to k_p$,  the Lorentz invariants $k_q\cdot k_a$ in the numerators and $s_{aq}$ in the denominators cancel each other, remaining the summation over $\delta_{aq}$, which vanishes due to the definition.

\subsection{General expansion of YMS amplitudes}
\label{subsecYSng}

In the previous subsection, the recursive expansion \eref{expand-recur-YS-2g} was observed from the sub-leading order soft behavior
of the YMS amplitude ${\cal A}_{YS}(1,\cdots,n;p,q|\sigma)$ provided in \eref{soft-A1}, while \eref{soft-A1} was obtained by acting the sub-leading soft operator on the YMS amplitude ${\cal A}_{YS}(1,\cdots,n;p|\sigma\setminus q)$. Such process suggests a recursive pattern, which leads to the general recursive expansion of YMS amplitudes, with arbitrary number of external gluons.

To give an example, we now derive the expansion of YMS amplitude ${\cal A}_{YS}(1,\cdots,n;p,q,r|\sigma)$ from the expansion of ${\cal A}_{YS}(1,\cdots,n;p,q|\sigma\setminus r)$, using the recursive pattern. We take $k_r\to \tau k_r$ and expand ${\cal A}_{YS}(1,\cdots,n;p,q,r|\sigma)$ in $\tau$, the sub-leading order contribution is determined by the soft theorem as
\bea
{\cal A}^{(1)}_{YS}(1,\cdots,n;p,q,r|\sigma)=S^{(1)}_g(r)\,{\cal A}_{YS}(1,\cdots,n;p,q|\sigma\setminus r)\,.~~~~\label{soft-theo-r}
\eea
Substituting the expansion of ${\cal A}_{YS}(1,\cdots,n;p,q|\sigma\setminus r)$ in \eref{expand-recur-YS-2g} into \eref{soft-theo-r}, we get
\bea
{\cal A}^{(1)}_{YS}(1,\cdots,n;p,q,r|\sigma)&=&(\epsilon_p\cdot Y_p)\,\Big[S^{(1)}_g(r)\,{\cal A}_{YS}(1,\{2,\cdots,n-1\}\shuffle p,n;q|\sigma\setminus r)\Big]\nn
& &+(\epsilon_p\cdot f_q\cdot X_q)\,\Big[S^{(1)}_g(r)\,{\cal A}_{S}(1,\{2,\cdots,n-1\}\shuffle \{q,p\},n|\sigma\setminus r)\Big]\nn
& &+\Big[S^{(1)}_g(r)\,(\epsilon_p\cdot Y_p)\Big]\,{\cal A}_{YS}(1,\{2,\cdots,n-1\}\shuffle p,n;q|\sigma\setminus r)\nn
& &+\Big[S^{(1)}_g(r)\,(\epsilon_p\cdot f_q\cdot X_q)\Big]\,{\cal A}_{S}(1,\{2,\cdots,n-1\}\shuffle \{q,p\},n|\sigma\setminus r)\,,~~~~\label{soft-theo-r-exp}
\eea
where the Leibnitz's rule has been used since the universal sub-leading soft operator for the gluon in \eref{soft-fac-g-1-1} includes the first order derivative of Lorentz vectors.
The first and second lines at the r.h.s of \eref{soft-theo-r-exp} can be recognized as
\bea
& &(\epsilon_p\cdot Y_p)\,\Big[S^{(1)}_g(r)\,{\cal A}_{YS}(1,\{2,\cdots,n-1\}\shuffle p,n;q|\sigma\setminus r)\Big]\nn
&=&(\epsilon_p\cdot Y_p)\,{\cal A}^{(1)}_{YS}(1,\{2,\cdots,n-1\}\shuffle p,n;q,r|\sigma)\,,~~~~\label{exp-1}
\eea
and
\bea
& &(\epsilon_p\cdot f_q\cdot X_q)\,\Big[S^{(1)}_g(r)\,{\cal A}_{S}(1,\{2,\cdots,n-1\}\shuffle \{q,p\},n|\sigma\setminus r)\Big]\nn
&=&
(\epsilon_p\cdot f_q\cdot X_q)\,{\cal A}^{(1)}_{YS}(1,\{2,\cdots,n-1\}\shuffle \{q,p\},n;r|\sigma)\,,~~~~\label{exp-2}
\eea
due to the soft theorem. The third line can be organized as
\bea
& &\Big[S^{(1)}_g(r)\,(\epsilon_p\cdot Y_p)\Big]\,{\cal A}_{YS}(1,\{2,\cdots,n-1\}\shuffle p,n;q|\sigma\setminus r)\nn
&=&\tau\,(\epsilon_p\cdot f_r\cdot Y_r)\,{\cal A}^{(0)}_{YS}(1,\{2,\cdots,n-1\}\shuffle \{r,p\},n;q|\sigma)\,,~~~~\label{exp-3}
\eea
via the manipulation similar to that in \eref{SeX} and \eref{SeXA}. Now we turn to the last line at the r.h.s of \eref{soft-theo-r-exp}.
Using the universal formula of $S^{(1)}_g(r)$ in \eref{soft-fac-g-1-1}, we have
\bea
S^{(1)}_g(r)\,(\epsilon_p\cdot f_q\cdot X_q)=H_1+H_2\,,
\eea
where
\bea
H_1&=&-{\delta_{pr}\,(\epsilon_p\cdot f_r\cdot f_q\cdot X_q)\over s_{pr}}
+{\delta_{qr}\,(\epsilon_p\cdot f_r\cdot f_q\cdot X_q)\over s_{qr}}\nn
H_2&=&-{\delta_{qr}\,(\epsilon_p\cdot f_q\cdot f_r\cdot X_q)\over s_{qr}}
+\sum_{i=1}^j\,{\delta_{ir}\,(\epsilon_p\cdot f_q\cdot f_r\cdot k_i)\over s_{ir}}\,,~~~~\label{H1H2}
\eea
with $X_q=\sum_{i=1}^j\,k_i$. Let us consider $H_1$ first. Based on the similar reason as that discussed below \eref{SeX},
we use $\delta_{ab}=-\delta_{ba}$ to reorganize $H_1$ which corresponds to the color ordering $(1,\cdots,i,q,i+1,\cdots,j,p,j+1,\cdots,n)$ as
follows
\bea
H_1=\Big[\Big({\delta_{qr}\over s_{qr}}+{\delta_{ri}\over s_{ri}}\Big)+\Big({\delta_{jr}\over s_{jr}}+{\delta_{rp}\over s_{rp}}\Big)+\sum_{k=i}^{j-1}\,\Big({\delta_{kr}\over s_{kr}}+{\delta_{r(k+1)}\over s_{r(k+1)}}\Big)\Big]\,(\epsilon_p\cdot f_r\cdot f_q\cdot X_q)\,.~~~~\label{H1}
\eea
Using \eref{H1} and the leading soft operator \eref{soft-fac-s-0} for the scalar, we find
\bea
& &H_1\,{\cal A}_S(1,\{2,\cdots,n-1\}\shuffle\{q,p\},n|\sigma\setminus r)\nn
&=&\tau\,(\epsilon_p\cdot f_r\cdot f_q\cdot X_q)\,{\cal A}^{(0)}_S(1,\{2,\cdots,n-1\}\shuffle\{q,r,p\},n|\sigma)\,.~~\label{ex-H1}
\eea
The form of $H_2$ in \eref{H1H2} is similar as the first line at the r.h.s of \eref{SeX}, with $\epsilon^\mu$ replaced by $(\epsilon\cdot f)^\mu$. Thus, one can perform the same manipulation as in \eref{SeX} and \eref{SeXA} to get
\bea
& &H_2\,{\cal A}_S(1,\{2,\cdots,n-1\}\shuffle\{q,p\},n|\sigma\setminus r)\nn
&=&\tau\,(\epsilon_p\cdot f_q\cdot f_r\cdot X_r)\,{\cal A}^{(0)}_S(1,\{2,\cdots,n-1\}\shuffle\{r,q,p\},n|\sigma)\,.~~\label{ex-H2}
\eea
Combining \eref{ex-H1} and \eref{ex-H2} together gives
\bea
& &\Big[S^{(1)}_g(r)\,(\epsilon_p\cdot f_q\cdot X_q)\Big]\,{\cal A}_S(1,\{2,\cdots,n-1\}\shuffle\{q,p\},n|\sigma\setminus r)\nn
&=&\tau\,(\epsilon_p\cdot f_r\cdot f_q\cdot X_q)\,{\cal A}^{(0)}_S(1,\{2,\cdots,n-1\}\shuffle\{q,r,p\},n|\sigma)\nn
& &+\tau\,(\epsilon_p\cdot f_q\cdot f_r\cdot X_r)\,{\cal A}^{(0)}_S(1,\{2,\cdots,n-1\}\shuffle\{r,q,p\},n|\sigma)\,.~~\label{exp-4}
\eea
Putting four pieces \eref{exp-1}, \eref{exp-2}, \eref{exp-3} and \eref{exp-4} together, we finally get
\bea
{\cal A}^{(1)}_{YS}(1,\cdots,n;p,q,r|\sigma)&=&(\epsilon_p\cdot Y_p)\,{\cal A}^{(1)}_{YS}(1,\{2,\cdots,n-1\}\shuffle p,n;q,r|\sigma)\nn
& &+(\epsilon_p\cdot f_q\cdot X_q)\,{\cal A}^{(1)}_{YS}(1,\{2,\cdots,n-1\}\shuffle \{q,p\},n;r|\sigma)\nn
& &+\tau\,(\epsilon_p\cdot f_r\cdot Y_r)\,{\cal A}^{(0)}_{YS}(1,\{2,\cdots,n-1\}\shuffle \{r,p\},n;q|\sigma)\nn
& &+\tau\,(\epsilon_p\cdot f_r\cdot f_q\cdot X_q)\,{\cal A}^{(0)}_S(1,\{2,\cdots,n-1\}\shuffle\{q,r,p\},n|\sigma)\nn
& &+\tau\,(\epsilon_p\cdot f_q\cdot f_r\cdot X_r)\,{\cal A}^{(0)}_S(1,\{2,\cdots,n-1\}\shuffle\{r,q,p\},n|\sigma)\,,~~~~\label{expand-3g-pre}
\eea
and subsequently
\bea
{\cal A}_{YS}(1,\cdots,n;p,q,r|\sigma)&=&(\epsilon_p\cdot Y_p)\,{\cal A}_{YS}(1,\{2,\cdots,n-1\}\shuffle p,n;q,r|\sigma)\nn
& &+(\epsilon_p\cdot f_q\cdot Y_q)\,{\cal A}_{YS}(1,\{2,\cdots,n-1\}\shuffle \{q,p\},n;r|\sigma)\nn
& &+(\epsilon_p\cdot f_r\cdot Y_r)\,{\cal A}_{YS}(1,\{2,\cdots,n-1\}\shuffle \{r,p\},n;q|\sigma)\nn
& &+(\epsilon_p\cdot f_r\cdot f_q\cdot Y_q)\,{\cal A}_S(1,\{2,\cdots,n-1\}\shuffle\{q,r,p\},n|\sigma)\nn
& &+(\epsilon_p\cdot f_q\cdot f_r\cdot Y_r)\,{\cal A}_S(1,\{2,\cdots,n-1\}\shuffle\{r,q,p\},n|\sigma)\,.~~~~\label{expand-recur-YS-3g}
\eea
In the expansion \eref{expand-recur-YS-3g}, we used $Y_a$ for $a=p,q,r$ in each coefficient, due to the following reason.
In \eref{expand-3g-pre}, replacing all $X_a$ by $Y_a$ yields no difference, since $X_q$ is equivalent to $Y_q$ in \eref{expand-recur-YS-2g}.
One can replace $X_q$ in \eref{expand-recur-YS-2g} by $Y_q$, and follow the procedure from \eref{soft-theo-r-exp} to \eref{expand-3g-pre},
to get the equivalent formula of \eref{expand-3g-pre}, with $X_a\to Y_a$. But $X_q$ is not equivalent to $Y_q$ in the second line at the r.h.s of \eref{expand-recur-YS-3g}.
Suppose we choose $X_q$ instead of $Y_q$ in this line, the sub-leading order contribution of such term is no longer the second line at the r.h.s of \eref{expand-3g-pre}, since for the cases $r$ is inserted at the l.h.s of $q$ in the color ordering, $X_q$ include $k_r$ which automatically carries $\tau$. Indeed, the $\tau k_r$ contributions in $X_q$ are collected into the fourth and fifth lines at the r.h.s of \eref{expand-3g-pre}, and correspond to fourth and fifth lines in \eref{expand-recur-YS-3g}. In the fourth and fifth lines in \eref{expand-recur-YS-3g}, $Y_a$ are equivalent to $X_a$.

The expansion of general YMS amplitudes with arbitrary number of external gluons can be achieved via the same recursive method, resulted in
\bea
{\cal A}_{YS}(1,\cdots,n;p_1\cdots,p_m|\sigma)=\sum_{\vec{\alpha}}\,C(\vec{\alpha})\,{\cal A}_{YS}(1,\{2,\cdots,n-1\}\shuffle\{\vec{\alpha},p_1\},n;\{p_2,\cdots,p_m\}\setminus \alpha|\sigma)\,,~~~~\label{expand-recur-YS-mg}
\eea
where each $\alpha=\{a_i\}$ with $i\in\{1,\cdots,k\},\,k\leq m-1$ is a subset of $\{p_2,\cdots,p_m\}$, and $\vec{\alpha}=\{a_1,\cdots,a_k\}$ is ordered. For $\alpha=\{p_2,\cdots,p_m\}$, the YMS amplitudes at the r.h.s are reduced to BAS ones. The summation is over all ordered sets $\vec{\alpha}$, rather than un-ordered $\alpha$. The coefficients $C(\vec{\alpha})$ are given as
\bea
C(\vec{\alpha})=\epsilon_{p_1}\cdot f_{a_k}\cdot f_{a_{k-1}}\cdots f_{a_1}\cdot Y_{a_1}\,.~~~~\label{c1}
\eea
To derive the general version \eref{expand-recur-YS-mg}, the following identities are useful,
\bea
\big(S^{(1)}_g(a)\,k_b\big)\cdot V=-{\delta_{ba}\over s_{ba}}\,(k_b\cdot f_a\cdot V)\,,~~~~\big(S^{(1)}_g(a)\,\epsilon_b\big)\cdot V=-{\delta_{ba}\over s_{ba}}\,(\epsilon_b\cdot f_a\cdot V)\,,~~~~\label{iden-1}
\eea
where $V$ is an arbitrary Lorentz vector, and
\bea
V_1\cdot\big(S^{(1)}_g(a)\,f_b\big)\cdot V_2={\delta_{ba}\over s_{ba}}\,V_1\cdot(f_a\cdot f_b-f_b\cdot f_a)\cdot V_2\,,~~~~\label{iden-2}
\eea
for two arbitrary Lorentz vectors $V_1$ and $V_2$. The above identities can be proved directly through the definition of
$S^{(1)}_g(a)$ in \eref{soft-fac-g-1-1}. The expansion of general YMS amplitudes to BAS basis can be obtained by substituting the recursive expansion
\eref{expand-recur-YS-mg} iteratively, as discussed in \cite{Fu:2017uzt}.

\section{Expanded YM amplitudes}
\label{secYM}

In this section, we derive the expansion of pure YM amplitudes, by applying the universal sub-leading soft operator for the gluon
given in \eref{soft-fac-g-1-1}. The method used in the previous section can not be applied to the YM case directly, since one can not take
the set of external scalars to be empty. To handle this difficult, in subsection.\ref{subsecrecur}, we develop another recursive pattern which generates the general
expansion of YMS amplitudes via the soft operator \eref{soft-fac-g-1-1}. Using the new recursive method, the expansion of YM amplitudes is determined in subsection.\ref{subsecexpandYM}.

\subsection{Another recursive pattern for YMS amplitudes}
\label{subsecrecur}

In this subsection, we discuss another recursive pattern, which generates the coefficients of YMS amplitudes in the recursive expansion
\eref{expand-recur-YS-mg}.

Suppose we know the first term in the recursive expansion \eref{expand-recur-YS-mg}, namely,
\bea
{\cal A}_{YS}(1,\cdots,n;p_1,\cdots,p_m|\sigma)=(\epsilon_{p_1}\cdot Y_{p_1})\,{\cal A}_{YS}(1,\{2,\cdots,n-1\}\shuffle p_1,n;p_2,\cdots,p_m|\sigma)+\cdots\,,~~~~\label{recur-1}
\eea
then the sub-leading order soft behavior of the external gluon $p_i$ with $i\in\{2,\cdots,m\}$ is given by
\bea
{\cal A}_{YS}^{(1)}(1,\cdots,n;p_1,\cdots,p_m|\sigma)&=&S^{(1)}_g(p_i)\,{\cal A}_{YS}(1,\cdots,n;\{p_1,\cdots,p_m\}\setminus p_i|\sigma\setminus p_i)\nn
&=&S^{(1)}_g(p_i)\,\Big[(\epsilon_{p_1}\cdot Y_{p_1})\,{\cal A}_{YS}(1,\{2,\cdots,n-1\}\shuffle p_1,n;\{p_2,\cdots,p_m\}\setminus p_i|\sigma\setminus p_i)\nn
& &+\cdots\Big]\,,~~~~\label{soft-impose}
\eea
due to the soft theorem. The second equality is obtained by substituting the expanded formula \eref{recur-1} into the first line at the r.h.s.
By applying the definition of $S^{(1)}_g(p_i)$ in \eref{soft-fac-g-1-1}, we have
\bea
& &{\cal A}_{YS}^{(1)}(1,\cdots,n;p_1,\cdots,p_m|\sigma)\nn
&=&(\epsilon_{p_1}\cdot Y_{p_1})\,\Big[S^{(1)}_g(p_i)\,{\cal A}_{YS}(1,\{2,\cdots,n-1\}\shuffle p_1,n;\{p_2,\cdots,p_m\}\setminus p_i|\sigma\setminus p_i)\Big]\nn
& &+\Big[S^{(1)}_g(p_i)\,(\epsilon_{p_1}\cdot Y_{p_1})\Big]\,{\cal A}_{YS}(1,\{2,\cdots,n-1\}\shuffle p_1,n;\{p_2,\cdots,p_m\}\setminus p_i|\sigma\setminus p_i)\nn
& &+\cdots\nn
&=&(\epsilon_{p_1}\cdot Y_{p_1})\,{\cal A}^{(1)}_{YS}(1,\{2,\cdots,n-1\}\shuffle p_1,n;p_2,\cdots,p_m|\sigma)\nn
& &+\tau\,(\epsilon_{p_1}\cdot f_{p_i}\cdot Y_{p_i})\,{\cal A}^{(0)}_{YS}(1,\{2,\cdots,n-1\}\shuffle \{p_i,p_1\},n;\{p_2,\cdots,p_m\}\setminus p_i|\sigma)\nn
& &+\cdots\,,~~~~\label{soft-recur-1}
\eea
where the second equality uses
\bea
& &{\cal A}^{(1)}_{YS}(1,\{2,\cdots,n-1\}\shuffle p_1,n;p_2,\cdots,p_m|\sigma)\nn
&=&S^{(1)}_g(p_i)\,{\cal A}_{YS}(1,\{2,\cdots,n-1\}\shuffle p_1,n;\{p_2,\cdots,p_m\}\setminus p_i|\sigma\setminus p_i)\,,
\eea
imposed by the soft theorem, as well as
\bea
& &\Big[S^{(1)}_g(p_i)\,(\epsilon_{p_1}\cdot Y_{p_1})\Big]\,{\cal A}_{YS}(1,\{2,\cdots,n-1\}\shuffle p_1,n;\{p_2,\cdots,p_m\}\setminus p_i|\sigma\setminus p_i)\nn
&=&\tau\,(\epsilon_{p_1}\cdot f_{p_i}\cdot Y_{p_i})\,{\cal A}^{(0)}_{YS}(1,\{2,\cdots,n-1\}\shuffle \{p_i,p_1\},n;\{p_2,\cdots,p_m\}\setminus p_i|\sigma)\,,
\eea
obtained by using the manipulation similar to that in \eref{SeX} and \eref{SeXA}. The formula
in \eref{soft-recur-1} indicates new terms in the expansion \eref{recur-1}, turns \eref{recur-1} to be
\bea
& &{\cal A}_{YS}(1,\cdots,n;p_1,\cdots,p_m|\sigma)\nn&=&(\epsilon_{p_1}\cdot Y_{p_1})\,{\cal A}_{YS}(1,\{2,\cdots,n-1\}\shuffle p_1,n;p_2,\cdots,p_m|\sigma)\nn
& &+\sum_{i\in\{2,\cdots,m\}}\,(\epsilon_{p_1}\cdot f_{p_i}\cdot Y_{p_i})\,{\cal A}_{YS}(1,\{2,\cdots,n-1\}\shuffle \{p_i,p_1\},n;\{p_2,\cdots,p_m\}\setminus p_i|\sigma)\nn
& &+\cdots\,.~~~~\label{recur-2}
\eea

Substituting the expanded formula in \eref{recur-2} into the first line at the r.h.s of \eref{soft-impose}, one see that the soft theorem imposes
\bea
& &{\cal A}_{YS}^{(1)}(1,\cdots,n;p_1,\cdots,p_m|\sigma)\nn
&=&S^{(1)}_g(p_i)\,\Big[(\epsilon_{p_1}\cdot Y_{p_1})\,{\cal A}_{YS}(1,\{2,\cdots,n-1\}\shuffle p_1,n;\{p_2,\cdots,p_m\}\setminus p_i|\sigma\setminus p_i)\nn
& &+\sum_{j\in\{2,\cdots,m\}\setminus i}\,(\epsilon_{p_1}\cdot f_{p_j}\cdot Y_{p_j})\,{\cal A}_{YS}(1,\{2,\cdots,n-1\}\shuffle \{p_j,p_1\},n;\{p_2,\cdots,p_m\}\setminus \{p_i,p_j\}|\sigma\setminus p_i)\nn
& &+\cdots\Big]\,.~~~~\label{soft-recur-2}
\eea
To continue the recursive process, we use the definition of
$S^{(1)}_g(p_i)$ to get
\bea
& &S^{(1)}_g(p_i)\,\Big[(\epsilon_{p_1}\cdot f_{p_j}\cdot Y_{p_j})\,{\cal A}_{YS}(1,\{2,\cdots,n-1\}\shuffle \{p_j,p_1\},n;\{p_2,\cdots,p_m\}\setminus \{p_i,p_j\}|\sigma\setminus p_i)\Big]\nn
&=&(\epsilon_{p_1}\cdot f_{p_j}\cdot Y_{p_j})\,\Big[S^{(1)}_g(p_i)\,{\cal A}_{YS}(1,\{2,\cdots,n-1\}\shuffle \{p_j,p_1\},n;\{p_2,\cdots,p_m\}\setminus \{p_i,p_j\}|\sigma\setminus p_i)\Big]\nn
& &+\Big[S^{(1)}_g(p_i)\,(\epsilon_{p_1}\cdot f_{p_j}\cdot Y_{p_j})\Big]\,{\cal A}_{YS}(1,\{2,\cdots,n-1\}\shuffle \{p_j,p_1\},n;\{p_2,\cdots,p_m\}\setminus \{p_i,p_j\}|\sigma\setminus p_i)\nn
&=&(\epsilon_{p_1}\cdot f_{p_j}\cdot Y_{p_j})\,{\cal A}^{(1)}_{YS}(1,\{2,\cdots,n-1\}\shuffle \{p_j,p_1\},n;\{p_2,\cdots,p_m\}\setminus p_j|\sigma)\nn
& &+\tau\,(\epsilon_{p_1}\cdot f_{p_i}\cdot f_{p_j}\cdot Y_{p_j})\,{\cal A}^{(0)}_{YS}(1,\{2,\cdots,n-1\}\shuffle \{p_j,p_i,p_1\},n;\{p_2,\cdots,p_m\}\setminus \{p_i,p_j\}|\sigma)\nn
& &+\tau\,(\epsilon_{p_1}\cdot f_{p_j}\cdot f_{p_i}\cdot Y_{p_i})\,{\cal A}^{(0)}_{YS}(1,\{2,\cdots,n-1\}\shuffle \{p_i,p_j,p_1\},n;\{p_2,\cdots,p_m\}\setminus \{p_i,p_j\}|\sigma)\,,
\eea
which adds new terms
\bea
& &\sum_{\substack{i,j\in\{2,\cdots,m\}\\i\neq j}}\,(\epsilon_{p_1}\cdot f_{p_i}\cdot f_{p_j}\cdot Y_{p_j})\,{\cal A}_{YS}(1,\{2,\cdots,n-1\}\shuffle \{p_j,p_i,p_1\},n;\{p_2,\cdots,p_m\}\setminus \{p_i,p_j\}|\sigma)\nn
& &+\sum_{\substack{i,j\in\{2,\cdots,m\}\\i\neq j}}\,(\epsilon_{p_1}\cdot f_{p_j}\cdot f_{p_i}\cdot Y_{p_i})\,{\cal A}_{YS}(1,\{2,\cdots,n-1\}\shuffle \{p_i,p_j,p_1\},n;\{p_2,\cdots,p_m\}\setminus \{p_i,p_j\}|\sigma)
\eea
to the expansion \eref{recur-2}.

Repeating the above procedure, one can arrive at the recursive expansion of YMS amplitudes given in \eref{expand-recur-YS-mg}.
The above method requires knowing the first term in the expansion which includes YMS amplitudes with $m-1$ external gluons, and generates the remaining terms from the first one.
For the YMS case, this method is not efficient, since it is not easy to obtain the first term. However, as will be seen in the next subsection,
for the pure YM case, the first term can be fixed via the simple argument, thus the above method yields the recursive expansion of
YM amplitudes to YMS ones.

\subsection{Expansion of YM amplitudes}
\label{subsecexpandYM}

This subsection devotes to derive the expansion of color ordered YM amplitudes by employing the recursive method described in the previous subsection.

Let us consider the $n$-point YM amplitude ${\cal A}_{Y}(\sigma)$ which carries the color ordering $\sigma$. For convenience, from now on we use $i\in\{1,\cdots,n\}$ to denote external gluons rather than scalars. To apply the recursive method introduced in the previous subsection, the first step is to find the first term in the expansion, which consists of YMS amplitudes with the minimum number of external scalars. Suppose ${\cal A}_{Y}(\sigma)$
can be expanded to YMS amplitudes, the minimum number of scalars carried by YMS amplitudes should be $2$. The reason is the YMS amplitude with only one external scalar does not exist, as can be seen from the Feynman rules, and can also be understood via the leading order soft operator for the scalar given in \eref{soft-fac-s-0}. For the YMS amplitude ${\cal A}_{YS}(p;1,\cdots,n|\sigma')$, where $p$ denotes the only external scalar and $\sigma'$ stands for the overall color ordering among all external legs, one can take the scalar $p$ to be the soft particle, to obtain a soft factor times the YM amplitude ${\cal A}_{Y}(\sigma'\setminus p)$. However, since the soft operator in \eref{soft-fac-s-0} only acts on external scalars, the above soft behavior is forbidden by the universality of the soft factor. This observation implies that
the YMS amplitude ${\cal A}_{YS}(p;1,\cdots,n|\sigma')$ can not exist. Thus, the first term in the recursive expansion for ${\cal A}_{Y}(\sigma)$ consists of YMS amplitudes with two external scalars. The KK relation shows that both the YM and YMS amplitudes can be expanded to BAS amplitudes ${\cal A}_S(1,\sigma_1,n;\sigma)$ with $1$ and $n$ fixed at two ends in the color ordering, where $\sigma_1$ denotes the permutation among $n-2$ legs in $\{2,\cdots,n-1\}$. It means the YMS amplitude contained in the first term can be fixed as ${\cal A}_{YS}(1,n;2,\cdots,n-1|\sigma)$.

Then we need to figure out the coefficient of ${\cal A}_{YS}(1,n;2,\cdots,n-1|\sigma)$. In ${\cal A}_{YS}(1,n;2,\cdots,n-1|\sigma)$, the coupling constants for all vertices are the coupling constant of YM theory, thus the mass dimension of ${\cal A}_{YS}(1,n;2,\cdots,n-1|\sigma)$ is the same as that of the YM amplitude ${\cal A}_{Y}(\sigma)$. Consequently, the coefficient of ${\cal A}_{YS}(1,n;2,\cdots,n-1|\sigma)$ has mass dimension $0$. On the other hand, the YM amplitude is linear in polarization vectors $\epsilon_1$
and $\epsilon_2$, but these polarization vectors are not included in ${\cal A}_{YS}(1,n;2,\cdots,n-1|\sigma)$. To summarize, the coefficient is a Lorentz invariant, with mass dimension $0$, linear in both $\epsilon_1$ and $\epsilon_n$, and does not contain any pole. There is only one candidate $\epsilon_1\cdot\epsilon_n$ which satisfies all of above requirements. The discussion mentioned above fixes the first term in the recursive expansion to be
\bea
(\epsilon_n\cdot\epsilon_1)\,{\cal A}_{YS}(1,n;2,\cdots,n-1|\sigma)\,,
\eea
namely,
\bea
{\cal A}_{Y}(\sigma)=(\epsilon_n\cdot\epsilon_1)\,{\cal A}_{YS}(1,n;2,\cdots,n-1|\sigma)+\cdots\,,~~\label{YM-start}
\eea

Now we can use the recursive method to figure out the remaining terms in \eref{YM-start}. Taking $k_i\to \tau k_i$ for one external gluon $i$
and expanding in $\tau$, the soft theorem gives
\bea
{\cal A}^{(1)}_{Y}(\sigma)&=&S^{(1)}_g(i)\,{\cal A}_{Y}(\sigma\setminus i)\nn
&=&S^{(1)}_g(i)\,\Big[(\epsilon_n\cdot\epsilon_1)\,{\cal A}_{YS}(1,n;\{2,\cdots,n-1\}\setminus i|\sigma\setminus i)+\cdots\Big]\,,
\eea
where we have substituted \eref{YM-start} into the first line at the r.h.s to get the second. By applying the definition of the soft operator
$S^{(1)}_g(i)$, we find
\bea
& &S^{(1)}_g(i)\,\Big[(\epsilon_n\cdot\epsilon_1)\,{\cal A}_{YS}(1,n;\{2,\cdots,n-1\}\setminus i|\sigma\setminus i)\Big]\nn
&=&(\epsilon_n\cdot\epsilon_1)\,\Big[S^{(1)}_g(i)\,{\cal A}_{YS}(1,n;\{2,\cdots,n-1\}\setminus i|\sigma\setminus i)\Big]+\Big[S^{(1)}_g(i)\,(\epsilon_n\cdot\epsilon_1)\Big]\,{\cal A}_{YS}(1,n;\{2,\cdots,n-1\}\setminus i|\sigma\setminus i)\nn
&=&(\epsilon_n\cdot\epsilon_1)\,{\cal A}^{(1)}_{YS}(1,n;2,\cdots,n-1|\sigma)+\tau\,(\epsilon_n\cdot f_i\cdot\epsilon_1)\,{\cal A}^{(0)}_{YS}(1,i,n;\{2,\cdots,n-1\}\setminus i|\sigma)\,,~~~\label{recur-YM-1}
\eea
where the soft theorem and the second identity in \eref{iden-1} have been used to get the last line. The result in \eref{recur-YM-1}
detects new terms in \eref{YM-start}, leads to
\bea
{\cal A}_{Y}(\sigma)&=&(\epsilon_n\cdot\epsilon_1)\,{\cal A}_{YS}(1,n;2,\cdots,n-1|\sigma)\nn
& &+\sum_{i\in\{2,\cdots,n-1\}}\,(\epsilon_n\cdot f_i\cdot\epsilon_1)\,{\cal A}_{YS}(1,i,n;\{2,\cdots,n-1\}\setminus i|\sigma)\,.~~\label{YM-recur-1}
\eea

One can continue the recursive process by substituting \eref{YM-recur-1} into the first line of \eref{recur-YM-1}, and computing
\bea
S^{(1)}_g(i)\,\Big[(\epsilon_n\cdot f_j\cdot\epsilon_1)\,{\cal A}_{YS}(1,j,n;\{2,\cdots,n-1\}\setminus \{i,j\}|\sigma\setminus i)\Big]\,.
\eea
Repeating the same procedure, the full recursive expansion for the YM amplitude is found to be
\bea
{\cal A}_{Y}(\sigma)=\sum_{\vec{\alpha}}\,\hat{C}(\vec{\alpha})\,{\cal A}_{YS}(1,a_1,\cdots,a_k,n;\{2,\cdots,n-1\}\setminus\alpha|\sigma)\,,~~~~\label{expand-recur-YM}
\eea
where $\alpha=\{a_i\}$ with $i\in\{1,\cdots,k\},\,k\leq n-2$ is a subset of $\{2,\cdots,n-1\}$, and $\vec{\alpha}=\{a_1,\cdots,a_k\}$ is ordered. The summation is over all ordered set $\vec{\alpha}$, and the coefficients $\hat{C}(\vec{\alpha})$ are
\bea
\hat{C}(\vec{\alpha})=\epsilon_n\cdot f_{a_k}\cdot f_{a_{k-1}}\cdots f_{a_1}\cdot\epsilon_1\,.~~\label{c2}
\eea
By substituting the recursive expansion of YMS amplitudes \eref{expand-recur-YS-mg} into \eref{expand-recur-YM} iteratively, one can get
the expansion of YM amplitudes to the BAS basis.

In the end of this subsection, we notice that for the pure YM amplitude ${\cal A}_{Y}(\sigma)$ which carries only one color ordering $\sigma$,
the soft factors \eref{soft-fac-g-0-1} and \eref{soft-fac-g-1-1} for the gluon are simplified to
\bea
S^{(0)}_g(a)&=&{1\over \tau}\,\Big({\epsilon_a\cdot k_{a-1}\over s_{a(a-1)}}-{\epsilon_a\cdot k_{a+1}\over s_{a(a+1)}}\Big)\,,\nn
S^{(1)}_g(a)&=&{\epsilon_a\cdot J_{a-1}\cdot k_a\over s_{a(a-1)}}-{\epsilon_a\cdot J_{a+1}\cdot k_a\over s_{a(a+1)}}\,,~~~\label{soft-fac-g-0-2}
\eea
which are the same as operators derived in \cite{Casali:2014xpa,Schwab:2014xua}.
Here $a-1$ and $a+1$ denote external legs adjacent to $a$ in the color ordering $\sigma$.

\section{Expanded EYM and GR amplitudes, and soft factors for graviton}
\label{secEYMGR}

In this section, we study the expansions of single trace EYM and GR amplitudes, as well as the soft factors for the graviton. In subsection.\ref{subsecexpandEYMGR}, we point out that the expansions of EYM and GR amplitudes to the KK YM basis can be generated from the expansions
of YMS and YM amplitudes directly, via the double copy structure. As an alternative method, we also use the soft theorem and the universality of soft factor to derive
the expansion of the EYM amplitude with one external graviton. In subsection.\ref{subsecsfoth}, we use the expanded EYM amplitude, and the soft theorem,
to determine the soft factors for the graviton, at leading, sub-leading, and sub-sub-leading orders.

\subsection{Expanded EYM and GR amplitudes}
\label{subsecexpandEYMGR}

In section.\ref{secYS} and section.\ref{secYM}, we found the recursive expansions for single trace YMS and pure YM amplitudes, provided in \eref{expand-recur-YS-mg} and \eref{expand-recur-YM}, respectively. Such recursive expansions can be generalized to single trace EYM and pure GR amplitudes directly, based on the double copy structure.
As discussed in subsection.\ref{subsecexpand}, the double copy indicates the EYM and GR amplitudes can be expanded to YM ones, with the coefficients the same as expanding the YMS and YM amplitudes to BAS ones. It means we have the analogous recursive expansions
\bea
{\cal A}_{EY}(1,\cdots,n;p_1\cdots,p_m)=\sum_{\vec{\alpha}}\,C(\vec{\alpha})\,{\cal A}_{EY}(1,\{2,\cdots,n-1\}\shuffle\{\vec{\alpha},p_1\},n;\{p_2,\cdots,p_m\}\setminus \alpha)\,,~~~~\label{expand-recur-EY-mh}
\eea
and
\bea
{\cal A}_{G}(1,\cdots,n)=\sum_{\vec{\alpha}}\,\hat{C}(\vec{\alpha})\,{\cal A}_{EY}(1,a_1,\cdots,a_k,n;\{2,\cdots,n-1\}\setminus\alpha)\,,~~~~\label{expand-recur-G}
\eea
with coefficients $C(\vec{\alpha})$ and $\hat{C}(\vec{\alpha})$ in \eref{c1} and \eref{c2} respectively. In the notation ${\cal A}_{EY}(\cdots;\cdots)$, legs at the l.h.s of $;$ are gluons, while those at the r.h.s are gravitons. The external gluons are color ordered, and the gravitons carry no color ordering.
The recursive expansions in \eref{expand-recur-EY-mh} and \eref{expand-recur-G} can also be derived via our method used in
section.\ref{secYS} and section.\ref{secYM}. However, since one can not conclude the existence of these recursive expansions without assuming the double copy structure, such derivation can not give new understanding than getting expansions from the double copy structure directly. Thus, in this subsection, we only give the simplest example, the derivation of the expansion for EYM amplitude ${\cal A}_{EY}(1,\cdots,n;p)$, which contains $n$ external gluons, and only one external graviton encoded by $p$. We also clarify that the soft factor for the gluon does not act on gravitons. The method in this subsection is extremely similar to that in subsection.\ref{subsecYS1g}.

Our plain is to express the expansion of ${\cal A}_{EY}(1,\cdots,n;p)$ as
\bea
{\cal A}_{EY}(1,\cdots,n;p)=\sum_{i=1}^{n-1}\,(\W\epsilon_p\cdot P_i)\,{\cal A}_{Y}(1,\cdots,i,p,i+1,\cdots,n)\,,~~~~\label{expand-EY-1h}
\eea
due to the similar reason as that for obtaining the formula \eref{expand-YS-1g}, and determine the combinatory momenta $P_i$ via the soft theorem.
Taking $k_1\to \tau k_1,$ and expanding ${\cal A}_{EY}(1,\cdots,n;p)$ in $\tau$,
the leading order contribution is given as
\bea
{\cal A}^{(0)}_{EY}(1,\cdots,n;p)=\sum_{i=1}^{n-1}\,(\W\epsilon_p\cdot P^{(0)}_i)\,{\cal A}^{(0)}_{Y}(1,\cdots,i,p,i+1,\cdots,n)\,,~~~~\label{soft-EY-1h-0}
\eea
where $P^{(0)}_i$ are again leading order contributions of $P_i$. Using the soft theorem and the soft factor in \eref{soft-fac-g-0-2},
we get
\bea
{\cal A}^{(0)}_{Y}(1,\cdots,i,p,i+1,\cdots,n)&=&{1\over \tau}\,\Big({\epsilon_1\cdot k_n\over s_{n1}}-{\epsilon_1\cdot k_2\over s_{12}}\Big)\,
{\cal A}_{Y}(\not{1},2,\cdots,i,p,i+1,\cdots,n)\,,~~{\rm for}~i\geq2\,,\nn
{\cal A}^{(0)}_{Y}(1,p,2,\cdots,n)&=&{1\over \tau}\,\Big({\epsilon_1\cdot k_n\over s_{n1}}-{\epsilon_1\cdot k_p\over s_{1p}}\Big)\,
{\cal A}_{Y}(\not{1},p,2,\cdots,n)\,,~~{\rm for}~i=1\,.~~~~\label{soft-g-0--}
\eea
Substituting \eref{soft-g-0--} into \eref{soft-EY-1h-0} gives
\bea
{\cal A}^{(0)}_{EY}(1,\cdots,n;p)&=&{1\over \tau}\,\Big({\epsilon_1\cdot k_n\over s_{n1}}-{\epsilon_1\cdot k_p\over s_{1p}}\Big)\,(\W\epsilon_p\cdot P^{(0)}_1)\,
{\cal A}_{Y}(\not{1},p,2,\cdots,n)\nn
& &+{1\over \tau}\,\Big({\epsilon_1\cdot k_n\over s_{n1}}-{\epsilon_1\cdot k_2\over s_{12}}\Big)\,\sum_{j=2}^{n-1}\,(\W\epsilon_p\cdot P^{(0)}_j)\,
{\cal A}_{Y}(\not{1},\cdots,j,p,j+1,\cdots,n)\,.~~~~\label{soft-EY-1g-0-2}
\eea
Since the soft theorem imposes
\bea
{\cal A}^{(0)}_{EY}(1,\cdots,n;p)=S^{(0)}_g(1)\,{\cal A}_{EY}(\not{1},\cdots,n;p)\,,~~~~\label{soft-theo-EY-1}
\eea
the universality of the soft operator $S^{(0)}_g(1)$ implies that the combinatory momentum $P_1^{(0)}$ accompanied by the pole $1/s_{1p}$ must vanish. Thus we find $P_1=k_1$, and conclude that the operator $S^{(0)}_g(a)$ in \eref{soft-fac-g-0-2} only acts on adjacent gluons.
By considering the sub-leading order soft behavior of the gluon $1$, one can figure out that the sub-leading soft factor $S^{(1)}_g(a)$ also only acts on adjacent gluons.
Similar as in the YMS case in subsection.\ref{subsecYS1g}, such phenomenon can be understood from the Feynman diagram point of view.

The vanishing of $P^{(0)}_1$ eliminates the first line at the r.h.s of \eref{soft-EY-1g-0-2}, thus the soft theorem \eref{soft-theo-EY-1} together with the soft operator \eref{soft-fac-g-0-2} require
\bea
& &S^{(0)}_g(1)\,{\cal A}_{EY}(\not{1},\cdots,n;p)\nn
&=&S^{(0)}_g(1)\,\sum_{j=2}^{n-1}\,(\W\epsilon_p\cdot P^{(0)}_j)\,
{\cal A}_{Y}(\not{1},\cdots,j,p,j+1,\cdots,n)\,,~~~~\label{eq-soft-EY}
\eea
which indicates the expansion
\bea
{\cal A}_{EY}(2,\cdots,n;p)=\sum_{j=2}^{n-1}\,(\W\epsilon_p\cdot P^{(0)}_j)\,
{\cal A}_{Y}(2,\cdots,j,p,j+1,\cdots,n)\,,~~~~\label{expand-2-EY}
\eea
The expansions in \eref{expand-2-EY} and \eref{expand-EY-1h} are totally the same,
up to a relabeling. Thus, the solution $P_1=k_1$ indicates $P^{(0)}_2=k_2$ in \eref{expand-2-EY}, therefore
\bea
P_2=k_2+\alpha k_1\,.~~~~\label{a-EY}
\eea
The parameter $\alpha$ can be fixed by considering the soft behavior of the the external gluon $2$.
After taking $k_2\to \tau k_2$ and expanding \eref{expand-EY-1h} in $\tau$, the leading order term is found to be
\bea
{\cal A}^{(0)}_{EY}(1,\cdots,n;p)&=&{1\over \tau}\,\Big({\epsilon_2\cdot k_p\over s_{p2}}-{\epsilon_2\cdot k_3\over s_{23}}\Big)\,(\W\epsilon_p\cdot P_1^{(0)})\,
{\cal A}_Y(1,p,\not{2},\cdots,n)\nn
& &+{1\over \tau}\,\Big({\epsilon_2\cdot k_1\over s_{12}}-{\epsilon_2\cdot k_p\over s_{2p}}\Big)\,(\W\epsilon_p\cdot P_2^{(0)})\,
{\cal A}_Y(1,\not{2},p,\cdots,n)\nn
& &+{1\over \tau}\,\Big({\epsilon_2\cdot k_1\over s_{12}}-{\epsilon_2\cdot k_3\over s_{23}}\Big)\,\sum_{i=3}^{n-1}\,(\W\epsilon_p\cdot P_i^{(0)})\,
{\cal A}_Y(1,\not{2},\cdots,i,p,i+1,\cdots,n)\,.~~~~\label{leading-k2-EY}
\eea
The soft theorem \eref{soft-theo-EY-1} and the universality of soft operator impose the constraint
\bea
{\cal A}^{(0)}_{EY}(1,\cdots,n;p)={1\over \tau}\,\Big({\epsilon_2\cdot k_1\over s_{12}}-{\epsilon_2\cdot k_3\over s_{23}}\Big)\,
{\cal A}_{EY}(1,\not{2},\cdots,n;p)\,.~~~\label{ee}
\eea
By applying the expansion \eref{expand-EY-1h} to ${\cal A}_{EY}(1,3,\cdots,n;p)$ in \eref{ee}, with the solution $P_1=k_1$, and comparing with \eref{leading-k2-EY}, one can get the following equation
\bea
\Big({\epsilon_2\cdot k_p\over s_{p2}}-{\epsilon_2\cdot k_3\over s_{23}}\Big)\,(\W\epsilon_p\cdot P_1^{(0)})
+\Big({\epsilon_2\cdot k_1\over s_{12}}-{\epsilon_2\cdot k_p\over s_{2p}}\Big)\,(\W\epsilon_p\cdot P_2^{(0)})
=\Big({\epsilon_2\cdot k_1\over s_{12}}-{\epsilon_2\cdot k_3\over s_{23}}\Big)\,(\W\epsilon_p\cdot k_1)\,.~~~~\label{eq-1-EY}
\eea
The solution to \eref{eq-1-EY}
is $\alpha=1$, thus we have $P_1=k_1,\,P_2=k_1+k_2$.

Taking the soft limit of other external gluons successively, and applying the same method, one can find
\bea
P_i=\sum_{j=1}^i\,k_j\,,~~~~\label{Pi-EY}
\eea
thus the EYM amplitude with one external graviton can be expanded as
\bea
{\cal A}_{EY}(1,\cdots,n;p)=(\W\epsilon_p\cdot X_p)\,{\cal A}_{Y}(1,\{2,\cdots,n-1\}\shuffle p,n)\,.~~~~\label{expand-EY-1h-2}
\eea
As can be seen, the whole process from \eref{expand-EY-1h} to \eref{expand-EY-1h-2} is paralleled to that from \eref{expand-YS-1g} to \eref{expand-YS-1g-2} in subsection.\ref{subsecYS1g}, with the replacement $\delta_{ij}\to\epsilon_j\cdot k_i$ which reflects the color-kinematic duality.

\subsection{Soft factors for graviton}
\label{subsecsfoth}

In this subsection, we determine the soft factors for the graviton at leading, sub-leading, and sub-sub-leading orders, using the recursive expansion of EY amplitudes in \eref{expand-recur-EY-mh} and the soft theorem. Let us consider the EY amplitude ${\cal A}_{EY}(1,\cdots,n;p,q)$ which contains $n$ external gluons, and two external gravitons labeled by $p$ and $q$. By applying the general recursive expansion \eref{expand-recur-EY-mh}, one can expand ${\cal A}_{EY}(1,\cdots,n;p,q)$ as
\bea
{\cal A}_{EY}(1,\cdots,n;p,q)&=&(\W\epsilon_p\cdot Y_p)\,{\cal A}_{EY}(1,\{2,\cdots,n-1\}\shuffle p,n;q)\nn
& &+(\W\epsilon_p\cdot {\W f}_q\cdot Y_q)\,{\cal A}_Y(1,\{2,\cdots,n-1\}\shuffle\{q,p\},n)\,,~~~\label{expand-EY-2h}
\eea
and use this formula to study the soft behavior of the external graviton $q$. Here the strength tensor $\W f_a$ is $\W f_a^{\mu\nu}\equiv k^\mu_a\W\epsilon^\nu_a-\W\epsilon_a^\mu k_a^\nu$. Since $\W\epsilon_p\cdot \W f_q\cdot Y_q$
in the second line at the r.h.s of \eref{expand-EY-2h} is proportional to $\tau$ under the re-scaling $k_q\to \tau k_q$, the leading order term of ${\cal A}_{EY}(1,\cdots,n;p,q)$ only arises from the first line.
One can use \eref{expand-recur-EY-mh} to expand ${\cal A}_{EY}(1,\{2,\cdots,n-1\}\shuffle p,n;q)$ in \eref{expand-EY-2h} further, and use the leading soft factor for the gluon given in \eref{soft-fac-g-0-2}, to get
\bea
& &{\cal A}^{(0)}_{EY}(1,\cdots,n;p,q)\nn
&=&(\W\epsilon_p\cdot Y_p)\,(\W\epsilon_q\cdot X_q)\,{\cal A}^{(0)}_Y(1,\{2,\cdots,n-1\}\shuffle p\shuffle q,n)\nn
&=&{1\over \tau}\,\sum_{i=1}^{n-1}\,\Big[\sum_{j\in\{1,\cdots,n-1\}\setminus i}\,\Big({\epsilon_q\cdot k_j\over s_{jq}}-{\epsilon_q\cdot k_{j+1}\over s_{q(j+1)}}\Big)\,(\W\epsilon_p\cdot Y_p)\,(\W\epsilon_q\cdot X_q)\,{\cal A}_Y(1,\cdots,j,\not{q},j+1,\cdots,i,p,i+1,\cdots,n)\nn
& &+\Big({\epsilon_q\cdot k_i\over s_{iq}}-{\epsilon_q\cdot k_p\over s_{qp}}\Big)\,(\W\epsilon_p\cdot Y_p)\,(\W\epsilon_q\cdot X_q)\,{\cal A}_Y(1,\cdots,i,\not{q},p,i+1,\cdots,n)\nn
& &+\Big({\epsilon_q\cdot k_p\over s_{pq}}-{\epsilon_q\cdot k_{i+1}\over s_{q(i+1)}}\Big)\,(\W\epsilon_p\cdot Y_p)\,(\W\epsilon_q\cdot X_q)\,{\cal A}_Y(1,\cdots,i,p,\not{q},i+1,\cdots,n)\Big]\nn
&=&{1\over \tau}\,\Big[{(\epsilon_1\cdot k_p)\,(\W\epsilon_q\cdot k_p)\over s_{pq}}-{(\epsilon_q\cdot k_n)\,(\W\epsilon_q\cdot k_p)\over s_{qn}}
+\sum_{j=1}^{n-1}\,\Big({(\epsilon_q\cdot k_j)\,(\W\epsilon_q\cdot k_j)\over s_{jq}}-{(\epsilon_q\cdot k_n)\,(\W\epsilon_q\cdot k_j)\over s_{qn}}\Big)\Big]\,
{\cal A}_{EY}(1,\cdots,n;p)\nn
&=&{1\over \tau}\,\Big({(\epsilon_q\cdot k_p)\,(\W\epsilon_q\cdot k_p)\over s_{pq}}+\sum_{j=1}^n\,{(\epsilon_q\cdot k_j)\,(\W\epsilon_q\cdot k_j)\over s_{jq}}\Big)\,{\cal A}_{EY}(1,\cdots,n;p)\,.~~~~\label{1order-EY2h-h}
\eea
The above manipulation is paralleled to that in \eref{1order-YS2g-g}, with the replacement $\delta_{aq}\to \epsilon_q\cdot k_a$.
Since the soft theorem requires
\bea
{\cal A}^{(0)}_{EY}(1,\cdots,n;p,q)=S^{(0)}_h(q)\,{\cal A}^{(0)}_{EY}(1,\cdots,n;p)\,,
\eea
from the last line of \eref{1order-EY2h-h} we observe that
\bea
S^{(0)}_h(q)={1\over \tau}\sum_{a}\,{(\epsilon_q\cdot k_a)\,(\W\epsilon_q\cdot k_a)\over s_{aq}}={1\over \tau}\sum_{a}\,{k_a\cdot \varepsilon_q\cdot k_a\over s_{aq}}\,,~~~~\label{soft-fac-h-0}
\eea
which is the same as the formula found by Weinberg in \cite{Weinberg}.
Here $\varepsilon_q^{\mu\nu}$ is the polarization tensor of the graviton, and the summation is over all external particles including both gluons and gravitons.

Then we turn to the sub-leading order. The expanded formula \eref{expand-EY-2h} indicates
\bea
{\cal A}^{(1)}_{EY}(1,\cdots,n;p,q)&=&(\W\epsilon_p\cdot Y_p)\,{\cal A}^{(1)}_{EY}(1,\{2,\cdots,n-1\}\shuffle p,n;q)\nn
& &+\tau\,(\W\epsilon_p\cdot {\W f}_q\cdot Y_q)\,{\cal A}^{(0)}_Y(1,\{2,\cdots,n-1\}\shuffle\{q,p\},n)\nn
&=&(\W\epsilon_p\cdot Y_p)\,\Big[S^{(1)}_h(q)\,{\cal A}_{Y}(1,\{2,\cdots,n-1\}\shuffle p,n)\Big]\nn
& &+\Big[\Big(\sum_a\,{(\epsilon_q\cdot k_a)\,(\W\epsilon_q\cdot J_a\cdot k_q)\over s_{aq}}\Big)\,(\W\epsilon_p\cdot Y_p)\Big]\,{\cal A}_Y(1,\{2\cdots,n-1\}\shuffle p,n)\,.~~~\label{eq-2o-1}
\eea
To obtain the second equality in \eref{eq-2o-1}, we have used the soft theorem, and the relation
\bea
& &\tau\,(\W\epsilon_p\cdot {\W f}_q\cdot Y_q)\,{\cal A}^{(0)}_Y(1,\{2,\cdots,n-1\}\shuffle\{q,p\},n)\nn
&=&\Big[\Big(\sum_a\,{(\epsilon_q\cdot k_a)\,(\W\epsilon_q\cdot J_a\cdot k_q)\over s_{aq}}\Big)\,(\W\epsilon_p\cdot Y_p)\Big]\,{\cal A}_Y(1,\{2\cdots,n-1\}\shuffle p,n)\,,~~~\label{eq-2o}
\eea
which comes from the computation paralleled to that in \eref{SeX} and \eref{SeXA}. In the second line of \eref{eq-2o}, the summation is over all external legs $a$, and the effective part is the summation over legs which contribute $k_a$ or $\W\epsilon_a$ to $(\W\epsilon_p\cdot Y_p)$,
since $(\W\epsilon_q\cdot J_a\cdot k_q)(\W\epsilon_p\cdot Y_p)$ vanishes otherwise.
From \eref{eq-2o-1}, one can observe that
\bea
{\cal A}^{(1)}_{EY}(1,\cdots,n;p,q)&=&S^{(1)}_h(q)\,{\cal A}_{EY}(1,\cdots,n;p)\nn
&=&S^{(1)}_h(q)\,\Big[(\W\epsilon_p\cdot Y_p)\,{\cal A}_Y(1,\{2\cdots,n-1\}\shuffle p,n)\Big]\nn
&=&(\W\epsilon_p\cdot Y_p)\,\Big[S^{(1)}_h(q)\,{\cal A}_Y(1,\{2\cdots,n-1\}\shuffle p,n)\Big]\nn
& &+\Big[S^{(1)}_h(q)\,(\W\epsilon_p\cdot Y_p)\Big]\,{\cal A}_Y(1,\{2\cdots,n-1\}\shuffle p,n)\,,~~~~\label{eq-2o-2}
\eea
where the sub-leading soft factor for the graviton is given as
\bea
S^{(1)}_h(q)=\sum_a\,{(\epsilon_q\cdot k_a)\,(\W\epsilon_q\cdot J_a\cdot k_q)\over s_{aq}}=\sum_a\,{k_a\cdot\varepsilon_q\cdot J_a\cdot k_q\over s_{aq}}\,,~~~~\label{soft-fac-h-1}
\eea
which is the same as that found in \cite{Cachazo:2014fwa,Schwab:2014xua,Afkhami-Jeddi:2014fia}.
The second equality in \eref{eq-2o-2} uses \eref{expand-recur-EY-mh} to expand ${\cal A}_{EY}(1,\cdots,n;p)$ further. The third uses Leibnitz's rule, since the operator \eref{soft-fac-h-1} includes the first order derivative of Lorentz vectors.

Finally, we consider the sub-sub-leading order. At this order, we have the analogue of \eref{eq-2o-1} which is
\bea
{\cal A}^{(2)}_{EY}(1,\cdots,n;p,q)&=&(\W\epsilon_p\cdot Y_p)\,{\cal A}^{(2)}_{EY}(1,\{2,\cdots,n-1\}\shuffle p,n;q)\nn
& &+\tau\,(\W\epsilon_p\cdot {\W f}_q\cdot Y_q)\,{\cal A}^{(1)}_Y(1,\{2,\cdots,n-1\}\shuffle\{q,p\},n)\nn
&=&(\W\epsilon_p\cdot Y_p)\,\Big[S^{(2)}_h(q)\,{\cal A}_{Y}(1,\{2,\cdots,n-1\}\shuffle p,n)\Big]\nn
& &+\tau\,(\W\epsilon_p\cdot {\W f}_q\cdot Y_q)\,\Big[{\sum_a}'\,\Big({\epsilon_p\cdot J_a\cdot k_q\over s_{aq}}-{\epsilon_p\cdot J_{a+1}\cdot k_q\over s_{(a+1)q}}\Big)\,{\cal A}_Y(1,\{2,\cdots,n-1\}\shuffle p,n)\Big]\,,~~~\label{eq-3o-1}
\eea
where we have used the soft theorem and the sub-leading soft operator \eref{soft-fac-g-0-2} for the gluon to get the second equality.
The summation $\sum'_a$ is over all legs at the l.h.s of $p$ in the color ordering.
To continue, we observe that the derivation of
$\eref{eq-2o}$ will not be altered when replacing $\epsilon_q\cdot k_a$ with the arbitrary operator ${\cal O}_{aq}$
determined by external leg $q$ and it's adjacent partner $a$.
Thus we can generalize \eref{eq-2o} to
\bea
& &\sum_a\,\Big[(\W\epsilon_q\cdot J_a\cdot k_q)\,(\W\epsilon_p\cdot Y_p)\Big]\,\Big[{{\cal O}_{aq}\over s_{aq}}\,{\cal A}_Y(1,\{2\cdots,n-1\}\shuffle p,n)\Big]\nn
&=&\,(\W\epsilon_p\cdot {\W f}_q\cdot Y_q)\,\Big[{\sum_a}'\,\Big({{\cal O}_{aq}\over s_{aq}}-{{\cal O}_{(a+1)q}\over s_{(a+1)q}}\Big)\,{\cal A}_Y(1,\{2,\cdots,n-1\}\shuffle p,n)\Big]\,,~~~~\label{gener}
\eea
where the summation in the first line is again over all external legs, and the summation in the second line is again over all legs at the l.h.s of $p$ in the color ordering. Substituting \eref{gener} into \eref{eq-3o-1} with ${\cal O}_{aq}=\epsilon_q\cdot J_a\cdot k_q$, we obtain the following expression for ${\cal A}^{(2)}_{EY}(1,\cdots,n;p,q)$,
\bea
{\cal A}^{(2)}_{EY}(1,\cdots,n;p,q)
&=&(\W\epsilon_p\cdot Y_p)\,\Big[S^{(2)}_h(q)\,{\cal A}_{Y}(1,\{2,\cdots,n-1\}\shuffle p,n)\Big]\nn
& &+\tau\,\sum_a\,\Big[(\W\epsilon_q\cdot J_a\cdot k_q)\,(\W\epsilon_p\cdot Y_p)\Big]\,\Big[{\epsilon_q\cdot J_a\cdot k_q\over s_{aq}}\,{\cal A}_Y(1,\{2\cdots,n-1\}\shuffle p,n)\Big]\,,~~~\label{eq-fin-1}
\eea
which indicates
\bea
{\cal A}^{(2)}_{EY}(1,\cdots,n;p,q)
&=&S^{(2)}_h(q)\,{\cal A}_{EY}(1,\cdots,n,p)\nn
&=&(\W\epsilon_p\cdot Y_p)\,\Big[S^{(2)}_h(q)\,{\cal A}_{Y}(1,\{2,\cdots,n-1\}\shuffle p,n)\Big]\nn
& &+\tau\,\sum_a\,\Big[(\W\epsilon_q\cdot J_a\cdot k_q)\,(\W\epsilon_p\cdot Y_p)\Big]\,\Big[{\epsilon_q\cdot J_a\cdot k_q\over s_{aq}}\,{\cal A}_Y(1,\{2\cdots,n-1\}\shuffle p,n)\Big]\nn
& &+\Big[S^{(2)}_h(q)\,(\W\epsilon_p\cdot Y_p)\Big]\,{\cal A}_Y(1,\{2\cdots,n-1\}\shuffle p,n)\,,~~~\label{eq-fin}
\eea
where the sub-sub-leading soft operator for the graviton is given by
\bea
S^{(2)}_h(q)={\tau\over 2}\,\sum_a\,{(\epsilon_q\cdot J_a\cdot k_q)\,(\W\epsilon_q\cdot J_a\cdot k_q)\over s_{aq}}=-{\tau\over2}\,\sum_a\,{k_q\cdot J_a\cdot \varepsilon_q\cdot J_a\cdot k_q\over s_{aq}}\,,~~~\label{soft-fac-h-2}
\eea
the same as the operator provided in \cite{Cachazo:2014fwa,Schwab:2014xua,Afkhami-Jeddi:2014fia}.
In the second equality of \eref{eq-fin}, the Leibnitz's rule and the observation $S^{(2)}_h(q)\,(\W\epsilon_p\cdot Y_p)=0$ have been used.
The factor $1/2$ arises from the observation that swapping $\epsilon_q\cdot J_a\cdot k_q$ and $\W\epsilon_q\cdot J_a\cdot k_q$ also leads to the second line at the r.h.s of \eref{eq-fin-1}, since for gravitons of Einstein gravity under consideration, two sets of polarization vectors $\{\epsilon_i\}$ and $\{\W\epsilon_i\}$ are the same.

From the formulas provided in \eref{soft-fac-h-0}, \eref{soft-fac-h-1}, and \eref{soft-fac-h-2}, one can see that soft operators for the graviton are manifestly gauge invariant at each order, they act on all external legs including both gravitons and gluons.

\section{Conclusion and discussion}
\label{secconclu}

In this paper, by imposing soft theorems, the universality of soft factors, and assuming the double copy structure, we constructed the expansions of single trace YMS tree amplitudes and pure YM tree amplitudes to KK BAS basis, and determined the soft factors including the leading factor for the BAS scalar, the leading and sub-leading factors for the gluon. Our method also leads to the expansions of single trace EYM tree amplitudes and pure GR tree amplitudes, as demonstrated in the simplest example in subsection.\ref{subsecexpandEYMGR}. Using the expanded formula of EYM amplitude, we reproduced the soft factors for the graviton at leading, sub-leading, as well as sub-sub-leading orders.

From our results, one can see that the soft factor for the BAS scalar only acts on external BAS scalars, the soft factors for the gluon
acts on external scalars and gluons, while the soft factors for the graviton acts on all external particles. These observations imply that the action of soft factors depend on the charges carried by external particles. BAS theory carries two gauge groups $G_1$ and $G_2$, YM theory carries only one gauge group $G_1$, while GR carries non of them. Correspondingly, the soft factor for the BAS scalar only acts on external particles which carry both $G_1$ and $G_2$ charges, the soft factors for the gluon only acts on external particles which carry the $G_1$ charges, and the soft factors for the graviton acts on all external particles which carry the gravitational charges.

An interesting question is, if the factorization in the manner of \eref{softtheo} holds at any order? Using the expansions of amplitudes, one can calculate the Laurent expansion of the amplitude in the soft parameter $\tau$ to any order. However, it does not mean the universal soft operators can be extracted at any order. This is a potential future direction. If the higher order operators exist, we need to work out the explicit formulas of them. And if they do not exist, we need to understand the reason.

The expansions of tree amplitudes can be extended to a wide range of theories, as discussed in \cite{Zhou:2019mbe}. Thus, another interesting future direction is to answer if the expansions of other theories, as well as the explicit formulas of soft factors for other particles, can be determined by imposing the soft theorems and the universality of soft factors.

\section*{Acknowledgments}

The author would thank Prof. Yijian Du, Chang Hu and Linghui Hou, for helpful discussions and valuable suggestions.


\end{document}